\documentclass[twocolumn,amsmath,amssymb]{revtex4-2}
\usepackage{graphicx}
\usepackage[compat=1.1.0]{tikz-feynman}
\usepackage{enumitem}
\usepackage{booktabs}
\usepackage[hidelinks]{hyperref}
\usepackage{xcolor}
\usepackage{comment}
\usepackage[T1]{fontenc}
\usepackage[utf8]{inputenc} 

\begin{document}

\title{Diagrammatic Monte Carlo for positron-molecule many-body theory}
\author{T. A. Scott, S. K. Gregg, and D. G. Green}
\email{d.green@qub.ac.uk}
\affiliation{School of Mathematics \& Physics, Queen's University Belfast, Belfast BT7 1NN, Northern Ireland, United Kingdom}
\date{June 1, 2026}
\begin{abstract}
A diagrammatic Monte Carlo evaluation of the ladder series contributions to the correlation potential (self energy) of a positron in the field of a molecule is presented. 
 The $GW$@TDHF, virtual-positronium ($T$-matrix), and positron–hole Goldstone ladder series contributions are stochastically sampled order-by-order within the Tamm–Dancoff approximation, which is exact for the latter two classes, with Cesàro–Riesz resummation used to extrapolate to infinite order. Gaussian bases are employed and Coulomb matrix elements are represented via density fitting, with the three centre integrals the largest arrays required to be stored in memory. 
The stochastic approach thus realizes a reduction in memory of the largest arrays required by a factor on the order of the number of molecular orbitals in the basis $N\sim$10$^2$--10$^3$ compared to the exact deterministic solution of Bethe-Salpeter equations [J. Hofierka, B. Cunningham, C. M. Rawlins, C. H. Patterson and D. G. Green, Nature {\bf 606}, {688} (2022)]. 
Benchmark results for lithium hydride show quantitative agreement with exact diagonalisation, notably demonstrating the successful stochastic summation of the virtual-positronium infinite electron-positron ladder series. 
\end{abstract}

\maketitle
The theoretical description of correlated interactions of electrons and positrons with atoms and molecules remains a challenging many-body problem. Positron-molecule interactions represent a particularly demanding case \cite{Hofierka2022}. 
Positrons are repelled by the nuclei, but strongly polarize the electron cloud and attract individual electrons at short range. The resulting correlation potential is a delicate balance of long-range polarization, screening, and the non-perturbative contribution of virtual-positronium formation, in which an electron tunnels to the positron and is temporarily captured by it. These correlations dominate low-energy scattering, enhance annihilation rates and $\gamma$ spectra, and can overcome static repulsion to produce positron binding and positronic bonding. 
Developing fundamental understanding of positron interactions with atoms and molecules is required to e.g., enable proper interpretation of important positron-based materials science diagnostics  \cite{Tuomisto2013,Hugenschmidt2016}, advance antimatter-based technologies (traps, high-energy resolution beams \cite{Danielson:2015} and positron emission tomography \cite{PETbook2,PETbook,Moskal2024,JPET2025}), understand positrons in space \cite{posfun,prantzos_511_2011,Fuller:2019,Flambaum:2021} and develop novel molecular spectroscopy \cite{RevModPhys.82.2557,Gribakin2017}.

\begin{figure}[t!]
    \centering
\includegraphics[width=0.45\textwidth]{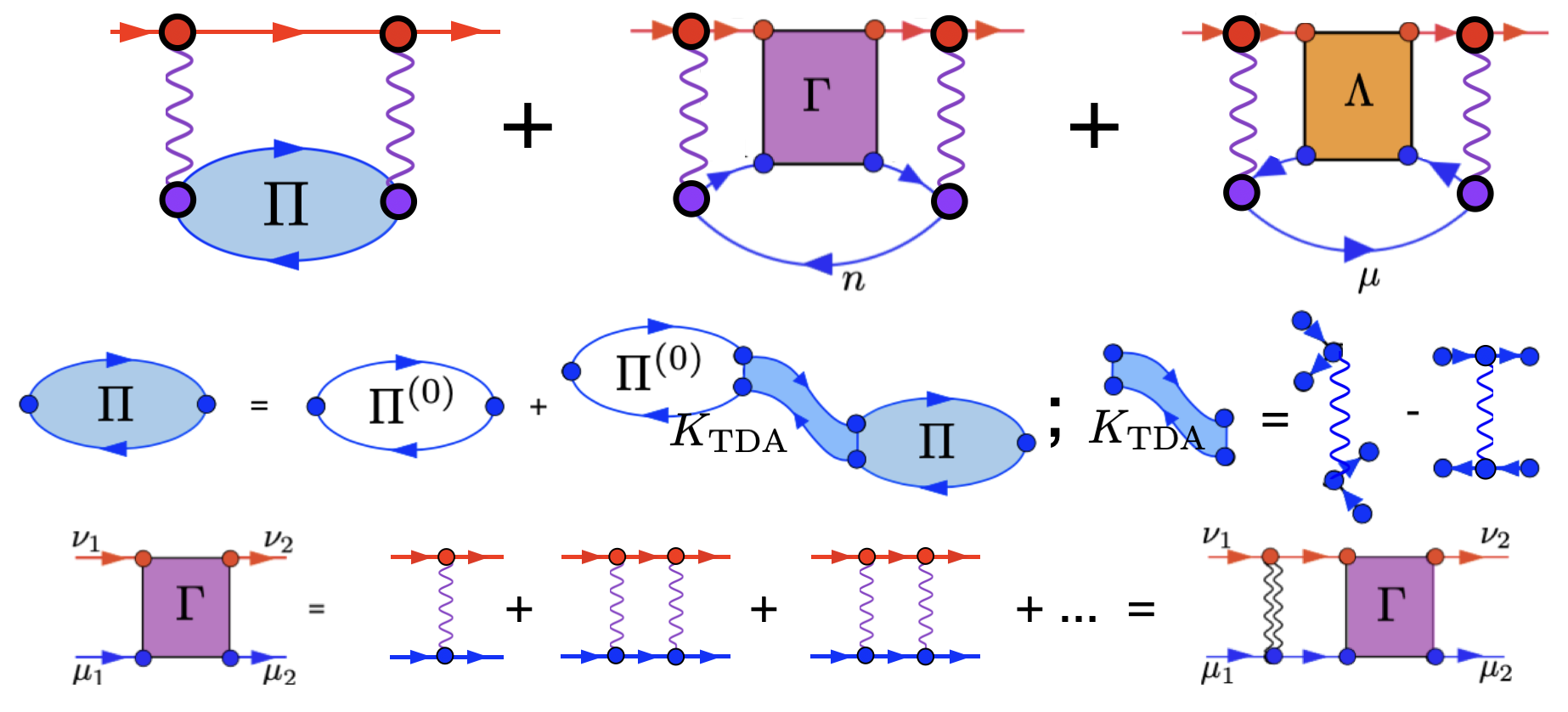}
    \caption{The three infinite classes of diagram contributions to the positron-molecule self energy summed here via diagMC: the $GW$@TDHF diagram, where $W=V\Pi V$ and $\Pi$ is the dressed electron-hole polarisation propagator in the TDA; the virtual-Ps electron-positron ladder series $\Gamma$; the analogous positron-hole ladder series $\Lambda$; and the Bethe-Salpeter equations for $\Pi$ and $\Gamma$. Red (blue) lines denote positron (electron/hole) HF propagators, purple (blue) wavy lines electron-positron (electron-electron) Coulomb interactions, on top of the $N$-electron ground-state molecule (white background).}
    \label{fig:sigma}
\end{figure}

Many-body theory is a powerful method that provides the natural and systematically improveable \emph{ab initio} account of correlations in atomic and molecular processes \cite{boylepindzola}, including positron-atom and molecule interactions \cite{Amusia:Pos:MBT:He,PhysScripta.46.248,PhysRevA.52.4541,Gribakin:2004,harabati2014identification,Cederbaum-elecpos,Green2013,Green:2014,Green2015,Rawlins2023}. 
In this approach, amplitudes of interest are calculated via infinite series of diagrams that involve products of Coulomb matrix elements and energy denominators summed over the intermediate states \cite{Fetter,mbtexposed,Hofierka2022}, but their evaluation is fundamentally constrained by the combinatorial growth of the number of diagram topologies, combinations of intermediate states \footnote{For example, for a virtual-Ps ladder contribution to the positron-molecule self energy with $n$ intermediate electron-positron interactions, i.e., at $(n+2)^{th}$ order in the electron-positron Coulomb interaction with spectator hole, one must sum over $N^{2(n+1)}$ combinations of the number of excited electron and positron intermediate states (MOs) in products of Coulomb matrix elements $\sum_{\mu_1\nu_1\dots \mu_{n+1},\nu_{n+1}}  \Pi_{k=1}^n V^k_{\mu_k\nu_k,\mu_{k+1}\nu_{k+1}} / \Delta^{k+1}_E$. Typically $N\gtrsim
  10^2$, thus the number of combinations grows beyond that capable for deterministic calculation swiftly.}, and large memory requirements of all-order resummations.
Specifically, considering positron-molecule interactions, the positron (quasiparticle) wavefunction of energy $\varepsilon$ in the field of the molecule is found via the Dyson equation \cite{Fetter,mbtexposed,Hofierka2022}  $\left(H_0+\Sigma_{\varepsilon}\right)\psi_\varepsilon=\varepsilon\psi_\varepsilon$, where $H_0$ is the zeroth-order Hamiltonian of the positron in the field of a target described in the Hartree-Fock approximation, and $\Sigma_E$ is the nonlocal, energy-dependent correlation potential (irreducible self energy) that is calculated diagrammatically. 
In our state-of-the-art implementation \cite{Hofierka2022}, $\Sigma$ contains several infinite classes of diagrams summed to all-orders via the solution of Bethe-Salpeter equations (BSE). These include the so-called $GW$ series \cite{Fetter} (which on its own is wholly deficient for the positron-molecule problem), the electron-positron ladder series that describes virtual-positronium formation, and the corresponding positron-hole series,  see \cite{Hofierka2022} and Fig.~\ref{fig:sigma} \footnote{For the positron-molecule problem the $GW$ diagram alone is wholly deficient. The importance of the virtual-positronium $\Gamma$ ladder series arises from the fact that successive terms in the series contribute with equal sign, in contrast to the all-electron case in which the signs alternate, leading to substantial cancellation. 
The virtual-positronium ladder series also corrects the annihilation amplitude vertex and its evaluation is crucial for accurate calculations of positron and positronium annihilation rates \cite{Green2015,Green2018,Swann2023}.}.
This deterministic all-order approach has enabled accurate calculations of positron binding energies and chemical insight for halogenated, ringed, and other polyatomic molecules in agreement with experiment \cite{Hofierka2022,Cassidy2024,Hofierka2024,ArthurBaidoo2024,Gregg2025,Gregg2026}
scattering cross sections and annihilation rates \cite{Rawlins2023,Hofierka2023,Gregg2025Gamma} and prediction of new types of positronically-bonded molecules \cite{Cassidy2024_2}. 
However, the price is severe: the BSE matrices inhabit large two-particle spaces and their solution via exact diagonalisation \cite{Shao:2016, Hofierka2022} requires a memory footprint of $\gtrsim8d_{\Gamma}^2$ bytes, where $d_{\Gamma} = N_\nu \times N_\mu$ is the product of the number of positron and excited-electron molecular orbitals (MOs) in the basis. Moreover, the need to simultaneously describe the short-range positron-repulsion from the nucleus, long-range polarisation, and virtual-Ps process that takes place away from but close to the molecule, necessitates larger bases than typical all-electron electronic structure  calculations. Typically, $N_{\nu}\sim N_{\mu}\sim $10$^2$--10$^3$, and thus the diagonalisation is extremely expensive, requiring communication-intensive dense MPI operations with a distributed-memory of $\lesssim$10 TB even for molecules with $\sim$10--20 atoms. A recent alternative coupled-cluster approach also highlighted the computational challenge of obtaining convergence \cite{Rosario2026}. 

The powerful alternative method of diagrammatic Monte Carlo (diagMC) (see e.g., \cite{Prokofev1998,VanHoucke2010,VanHoucke2012,Chen2019,Fedor2021,Azadi2022,Bighin2023,John2024,Vanhoecke2024,Stefano2025,Luo2025}) enables summation of Feynman diagrammatic series via stochastic sampling of the diagram order, internal quantum numbers, and topologies, allowing controllable resummation of infinite series \cite{VanHoucke2012}. It has provided spectacular success for lattice systems, polarons etc, and models of nuclei \cite{Brolli} but its successful application to particle interactions with real atoms and molecules has not yet been demonstrated.

In this work, we present a diagMC calculation for the self energy of a positron interacting with real molecules, importantly demonstrating its ability to converge infinite ladder series for such systems. Strikingly, we also show that it removes the memory bottleneck of the deterministic BSE approach, reducing the memory required by a factor of the number of MOs in the basis $N\sim10^2$--$10^3$.
 We consider the expansion of the self energy described via Goldstone diagrams in the orbital basis $\{|i\rangle;~H_0|i\rangle=\varepsilon_i|i\rangle$\}
in powers of the electron-positron and electron-electron Coulomb interactions $\Sigma_{if}(E)=\sum_{n=2}^{\infty} \Sigma_{if}^{(n)}(E)$,
where 
$\Sigma^{(n)}
=\sum_{\alpha\zeta}\mathcal{D}^{(n)}_{\alpha\zeta}(E,i,f)$ 
is the $n$th-order contribution and
$\mathcal{D}^{(n)}_{\alpha\zeta}$ is the weight of an individual diagram (summand) labelled by the internal positron, electron, and hole molecular-orbital indices denoted collectively as $\alpha$, and the toplogy $\zeta$. 
We aim for the stochastic evaluation of the series, and specifically, proof of principle in the ability for diagMC to calculate the non-perturbative virtual-Ps ladder series. 
By combining stochastic sampling of the $GW$@TDHF series in the Tamm-Dancoff approximation (TDA), the virtual-positronium ladder series, and positron-hole ladder series contributions to the self-energy, employing density-fitted Coulomb interactions  and Cesàro--Riesz resummation, we evaluate infinite-order many-body contributions without constructing the large two-particle Hamiltonians required by the exact diagonalisation solution of the Bethe-Salpeter equations.
The maximum array sizes involved are those of the three-centre density fitting integrals, $N_{\nu}^2N_{\rm aux}\sim 3N_{\nu}^3$, where $N_{\rm aux}$ is the size of the auxiliary density-fitting basis. 
Thus, the stochastic approach realises a striking reduction of memory by a factor of around $N\sim$10$^2$--10$^3$ compared to the exact diagonalisation approach. 
We benchmark the method by calculating positron binding energies in LiH, finding excellent agreement with the exact diagonalisation, extending the list of systems for which diagMC has been successfully applied to positron-molecule interactions, paving the way for its application to more general particle interactions with real atoms and molecules. 

\emph{Diagrammatic Monte Carlo evaluation of the positron-molecule self energy ladder series}.---
The individual matrix elements of the positron self energy matrix $\Sigma_{if}(E)$ are stochastically evaluated independently and in parallel as follows
\footnote{Since the self-energy matrix is symmetric, only the upper triangle ($i \leq f$) is sampled, with the calculation parallelised over unique $(i,f)$ pairs.}.
We first pre-compute electron-positron and electron-electron Coulomb matrix elements in the HF molecular bases in density-fitted three-centre integral form \cite{DF1,Hofierka2022}.
Exploiting the symmetry of the three-centre integrals, and saving in single-precision floating-point format, reduces memory by approximately 75\% relative to the full three-centre tensor in double precision, with negligible effect on the accuracy compared to the stochastic uncertainty of the Monte Carlo sampling.
The diagMC algorithm generates a Markov chain of diagram configurations using the Metropolis--Hastings algorithm \cite{Prokofev1998}, here with configurations sampled with probability proportional to the absolute value of their diagrammatic weight  $p(\mathcal{D})\propto|\mathcal{D}|$.
Specifically, a proposed diagram update $\mathcal{D}_a \to \mathcal{D}_b$ is accepted with probability
$    P^{\mathrm{accept}}_{a\to b} = \min\left(1,\; R_{a\to b}\right),$
where the acceptance ratio is
$    R_{a\to b} = \left|{\mathcal{D}_b/\mathcal{D}_a}\right|\,
    {P^{\mathrm{propose}}_{b\to a}}/{P^{\mathrm{propose}}_{a\to b}}$; we use uniform proposal probabilities. 
The update set is designed to satisfy detailed balance and ergodically explore the specified configuration space.
Since each Markov chain yields a self-energy element only up to an overall normalisation,  an auxiliary ``type-0'' sector is introduced with a known, strictly positive weight $\mathcal{D}_0$ that does not contribute to the physical series.
The ratio of type-0 visits $Z_0$ to total MC steps $N$  determines the normalisation constant:
${Z_0}/{N} = \mathcal{D}_0/{C}$,
where $C$ is the (unknown) total partition function.
The self-energy estimator then takes the form
\begin{equation}
    \label{eq:estimator}
    {\Sigma}_{if}(E) = \frac{\mathcal{D}_0}{Z_0}\,
    \sum_{k=1}^{N} \mathrm{sgn}\,\mathcal{D}_k(E,i,f),
\end{equation}
where $\mathrm{sgn}\,\mathcal{D}_k$ is the sign of the $k$-th sampled diagram. 
The choice of $\mathcal{D}_0$ must produce a reasonable balance between physical and unphysical samples across all $(i,f)$ pairs, whose self-energy magnitudes can span many orders of magnitude.
We set $\mathcal{D}_0$ equal to the absolute value of the second-order self-energy $\Sigma^{(2)}_{if}$, computed exactly.  
This choice is inexpensive to calculate for each $(if)$ pair, provides an internal consistency check against the Monte Carlo estimate at second order, and generally yields balanced sampling of physical and unphysical sectors. 
Our present algorithm employs three types of updates: 

1.~\emph{Transitions to and from the unphysical sector}.---A transition into the unphysical sector is proposed only when the current diagram is second order.
The diagram weight is replaced by $\mathcal{D}_0$ and the diagram is marked as unphysical.
The reverse move returns to the physical sector by constructing a second-order diagram with uniformly sampled internal indices $\nu$, $\mu$, and $n$.

2.~\emph{Adding and removing an interaction}.---The add update takes an $N$th-order diagram to order $N+1$ by inserting a new interaction vertex immediately before the final one.
The type of interaction is proposed with specified probability \footnote{The proposal probabilities are user-configurable. For example, selecting only the positron--electron interaction with all others set to zero yields the $\Gamma$ ladder series. When multiple interaction types are active, their proposal probabilities must be equal to satisfy detailed balance.}.
For example, the update from second-order to third-order 
inserts a new interaction between the first and last vertices:
\begin{center}
    \begin{tikzpicture}
        \begin{feynman}
            \vertex [draw, rectangle, fill=gray, minimum width=1cm, minimum height=1.5cm] (block) {$V$};
            \vertex [left=of block.north west] (2);
            \vertex [right=of block.north east] (3);
            \vertex [left=of 2] (1) {$i$};
            \vertex [right=of 3] (4) {$f$};
            \vertex [left=of block.south west] (elc2);
            \vertex [right=of block.south east] (elc3);
            \diagram*{
                (1) -- [fermion] (2) -- [fermion, edge label={$\nu_1$}] (block.north west);
                (block.north east) -- [fermion, edge label={$\nu_2$}] (3) -- [fermion] (4);
                (elc2) -- [fermion, quarter left, edge label={$\mu_1$}] (block.west);
                (block.east) -- [fermion, quarter left, edge label={$\mu_2$}] (elc3);
                (elc3) -- [fermion, quarter left, edge label={$n_2$}] (block.south east);
                (block.south west) -- [fermion, quarter left, edge label={$n_1$}] (elc2);

                (2) -- [photon] (elc2);
                (3) -- [photon] (elc3);
            };
        \end{feynman}
    \end{tikzpicture}
\end{center}
For a \emph{positron--electron} interaction, the $V$ block takes the form
\vspace*{-1ex}
\begin{center}
    \begin{tikzpicture}
        \begin{feynman}
            \vertex (2);
            \vertex [below=of 2] (elc2);
            \vertex [left=of 2] (1);
            \vertex [right=of 2] (3);
            \vertex [left=of elc2] (elc1);
            \vertex [right=of elc2] (elc3);

            \diagram*{
                (1) -- [fermion, edge label={$\nu_1$}] (2) -- [fermion, edge label={$\nu_2$}] (3);
                (elc1) -- [fermion, edge label={$\mu_1$}] (elc2) -- [fermion, edge label={$\mu_2$}] (elc3);

                (2) -- [photon] (elc2);
            };
        \end{feynman}
    \end{tikzpicture}
\end{center}
with vertex factor 
$    V_\mathrm{int} = -(\nu_1\nu_2|\mu_1\mu_2)$, and
    $n_2 = n_1.$
For a \emph{positron--hole} interaction,
\vspace*{-1ex}
\begin{center}
    \begin{tikzpicture}
        \begin{feynman}
            \vertex (2);
            \vertex [below=of 2] (elc2);
            \vertex [left=of 2] (1);
            \vertex [right=of 2] (3);
            \vertex [left=of elc2] (elc1);
            \vertex [right=of elc2] (elc3);

            \diagram*{
                (1) -- [fermion, edge label={$\nu_1$}] (2) -- [fermion, edge label={$\nu_2$}] (3);
                (elc3) -- [fermion, edge label={$n_2$}] (elc2) -- [fermion, edge label={$n_1$}] (elc1);

                (2) -- [photon] (elc2);
            };
        \end{feynman}
    \end{tikzpicture}
\end{center}
with
$    V_\mathrm{int} = (\nu_1\nu_2|n_1 n_2)$,
and    $\mu_2 = \mu_1.$
For \emph{electron--hole}, both direct and exchange interactions contribute:
\begin{center}
~~
    \begin{tikzpicture}
        \begin{feynman}
            \vertex(2);
            \vertex [below=0.75cm of 2] (elc1);
            \vertex [above left=0.5cm of 2, xshift=-0.8cm] (11) {$\mu_1$};
            \vertex [below left=0.5cm of 2, xshift=-0.8cm] (12) {$n_1$};
            \vertex [above right=0.5cm of elc2, xshift=0.8cm] (elc11) {$\mu_2$};
            \vertex [below right=0.5cm of elc2, xshift=0.8cm] (elc12) {$n_2$};

            \diagram*{
                (11) -- [fermion, quarter left] (2);
                (2) -- [fermion, quarter left] (12);
                (elc2) -- [fermion, quarter left] (elc11);
                (elc12) -- [fermion, quarter left] (elc2);

                (2) -- [photon] (elc2);
            };
        \end{feynman}
    \end{tikzpicture}
    \hspace{0.2cm}
    \begin{tikzpicture}
        \begin{feynman}
            \vertex (2);
            \vertex [below=of 2] (elc2);
            \vertex [left=of 2] (1);
            \vertex [right=of 2] (3);
            \vertex [left=of elc2] (elc1);
            \vertex [right=of elc2] (elc3);

            \diagram*{
                (1) -- [fermion, edge label={$\mu_1$}] (2) -- [fermion, edge label={$\mu_2$}] (3);
                (elc3) -- [fermion, edge label={$n_2$}] (elc2) -- [fermion, edge label={$n_1$}] (elc1);

                (2) -- [photon] (elc2);
            };
        \end{feynman}
    \end{tikzpicture}
\end{center}
Since both interactions share the same incoming and outgoing lines, both can be included in a single update:
    $V_\mathrm{int} = 2(\mu_1 n_1|n_2\mu_2) - (\mu_1\mu_2|n_1 n_2)$,
    with $\nu_2 = \nu_1$.
They give rise to the RPA and TDHF ladder series.

The new vertex introduces propagator lines whose MO indices are chosen uniformly at random.
The diagram weight is updated incrementally:
\begin{eqnarray}
\label{eq:sigma_n_to_n1}
    &&\mathcal{D}_{if}^{(N+1)} =\nonumber \\ &&\frac{\mathcal{D}_{if}^{(N)}}{(\nu_N f|\mu_N n_N)}\;
    \frac{V_\mathrm{int}\,(\nu_{N+1} f|n_{N+1}\mu_{N+1})}
    {E + \varepsilon_{n_{N+1}}- \varepsilon_{\nu_{N+1}} - \varepsilon_{\mu_{N+1}} },
\end{eqnarray}
requiring evaluation of only the modified factors, not the full diagram weight.
The reverse move removes the last internal interaction by applying the inverse transformation:
\begin{eqnarray}
    \label{eq:sigma_n_to_n1_rev}
    \mathcal{D}_{if}^{(N-1)} &=& (\nu_{N-1} f|n_{N-1}\mu_{N-1})\nonumber\\&\times& \frac{\mathcal{D}_{if}^{(N)}\,(E + \varepsilon_{n_N} - \varepsilon_{\nu_N} - \varepsilon_{\mu_N} )}
    {V_\mathrm{int}\,(\nu_N f|\mu_N n_N)}\;
    .
\end{eqnarray}

3.~\emph{Modifying an internal propagator line}.---The line-modification update selects a random internal propagator line (positron, electron, or hole), identifies the bounding interactions, and chooses a new MO index uniformly.
If $V_1$, $V_2$ and $V_1'$, $V_2'$ denote the old and new interactions bounding the selected line, and $\Delta E_k$, $\Delta E_k'$ denote the old and new energy denominators, the updated weight is
\begin{equation}
    \label{eq:update_modify}
    \mathcal{D}'_{if} = \mathcal{D}_{if}\;
    \frac{V_1'\,V_2'}{V_1\,V_2}
    \prod_{k}\frac{\Delta E_k}{\Delta E_k'}.
\end{equation}

In practice we accumulate self-energy matrix elements order-by-order. At each diagram order $n$, the $n$th-order contribution is reconstructed as
$\Sigma^{(n)}_{if} = \,A^{(n)}_{if}\mathcal{D}_0/{Z_0}$,
where $A^{(n)}_{if}$ is the accumulated signs and $Z_0$ is the number of type-0 visits, both specific to the $(i,f)$ pair.
The full self-energy up to order $N$ is obtained by the summation,
\begin{equation}
    \Sigma^{(\leq N)}(E) = \sum_{n=2}^{N} \Sigma^{(n)}(E).
\end{equation}

The partial sums $\Sigma^{(\leq N)}$ may converge slowly, oscillate, or diverge as $N$ increases, particularly for the $\Gamma$ series where strong positron--electron correlations lead to growth of diagram weights with increasing order.
To extract physically meaningful infinite-order estimates, we employ Ces\`aro--Riesz resummation~\cite{Korle1970}, which replaces the partial sum by the weighted sum
\begin{equation}
    \label{eq:cesaro}
    \Sigma^{(\leq N)}_\delta(E) = \sum_{n=2}^{N} \Sigma^{(n)}(E)\,F_N^{(n)},
    ~~
    F_N^{(n)} = \left(\frac{N - n + 1}{N}\right)^{\!\delta},
\end{equation}
where $\delta > 0$ controls the suppression of high-order terms.
For $n \ll N$, the factor $F_N^{(n)} \approx 1$, while for $n \approx N$ it is strongly damped.
In the limit $N \to \infty$, the resummation reproduces the original series.
The Dyson equation is then solved using the resummed self-energy matrix yielding the positron binding energy $\varepsilon_b$ (in practice we calculate over a grid of $E$ and interpolate to the self-consistent solution $\varepsilon_b=E$). The resulting binding energies are studied as a function of $1/N$ and extrapolated to $1/N \to 0$ by fitting the model $\varepsilon_b(1/N) = A\bigl(e^{B/N} - 1\bigr) + C,$
where $C$ gives the extrapolated binding energy.
At large $N$ the damping factors $F_N^{(n)}$ for fixed low-order $n$ deviate from unity by corrections of order $1/N$, so the resummed binding energy is a smooth function of $1/N$ whose leading correction is linear.
The exponential form captures this linear behaviour for large $N$ while accommodating nonlinear corrections at higher $N$; in practice it provides a more stable extrapolation than a polynomial fit.
We also found that restricting the fit to $N \geq 5$ (i.e.\ $1/N \leq 0.2$) improves the stability of the extrapolation.
To quantify the uncertainty in the extrapolation, we compute binding energies for $\delta$ ranging from $1$ to $3$ in steps of $0.1$, fit each to the model, and take the mean extrapolated value as the binding energy and the standard deviation as the error.

\emph{Results}.---We benchmark the approach against our exact diagonalisation approach for lithium hydride. Binding energies for LiH were previously calculated via the exact diagonalisation at the $GW$@BSE+$\Gamma+\Lambda$ levels including screened ladders by one of us and colleagues in \cite{Hofierka2022}. In the present work we calculate the $GW$ series is in the less sophisticated TDA and use bare Coulomb interactions, aiming here for a proof-of-principle that the diagMC approach can successfully calculate the virtual-positronium ladder series.
We use a Gaussian basis with standard aug-cc-pVQZ sets on the H Li, and five `ghost' centres surrounding the negative (H) end of the molecule, around $\sim$ 1 a.u. away from it. This results in $\sim$ 224, 271 and 518 electron, positron and auxillary MOs respectively; exact diagonalisation requires 29\,GB RAM, compared with the drastically reduced 0.1\,GB of diagMC. 

\begin{figure}[!ht]
    \centering
    \includegraphics[width=0.48\textwidth]{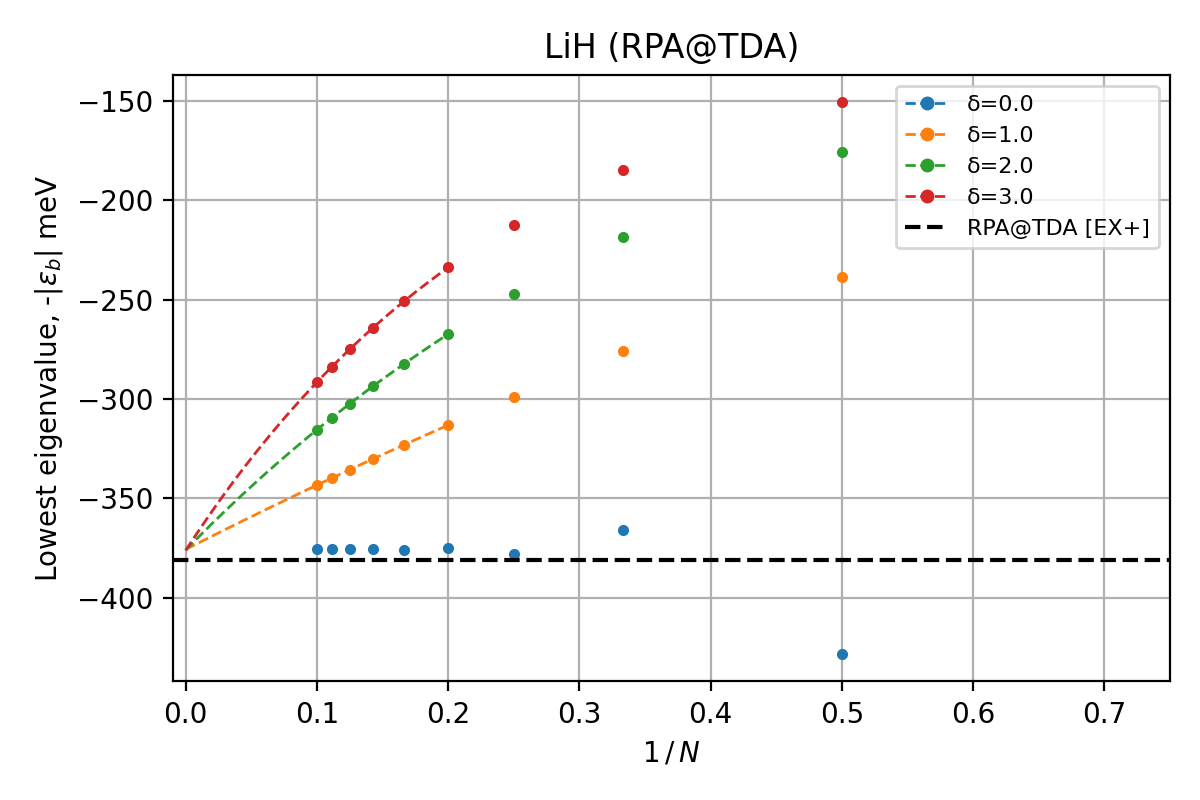}\\
    \includegraphics[width=0.48\textwidth]{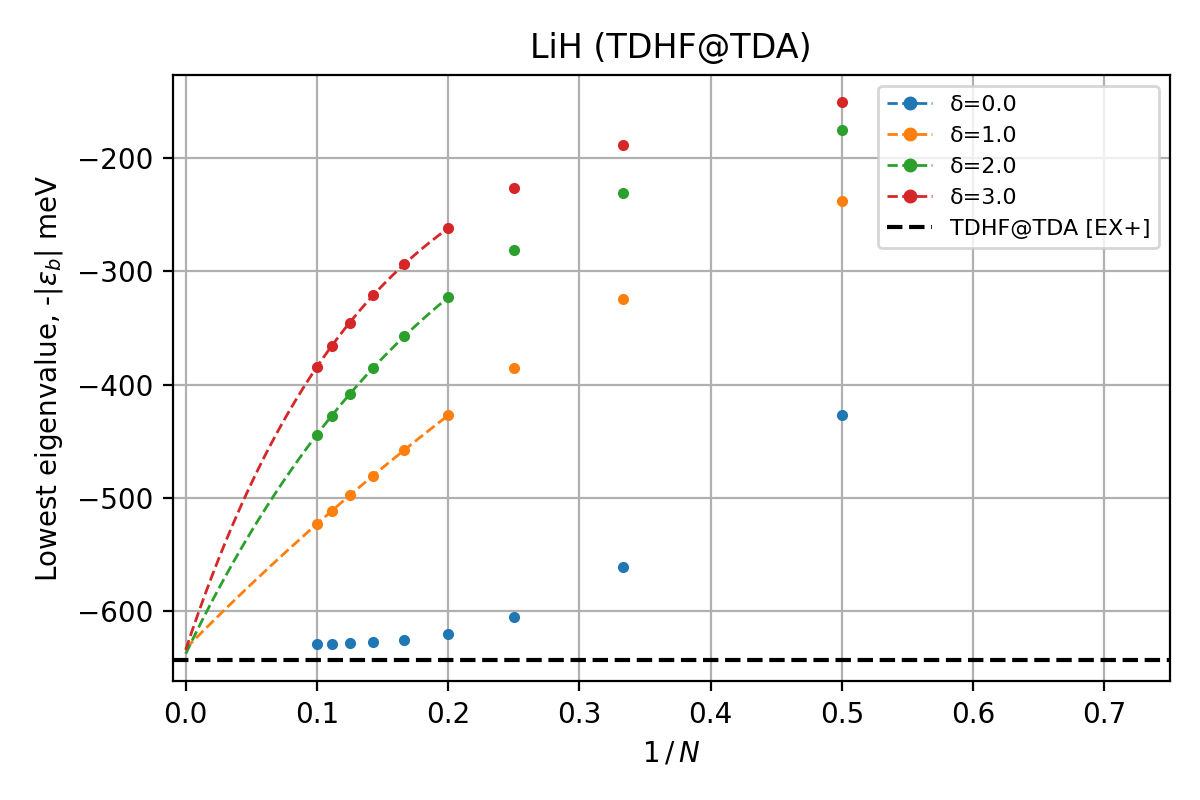}
    \includegraphics[width=0.48\textwidth]{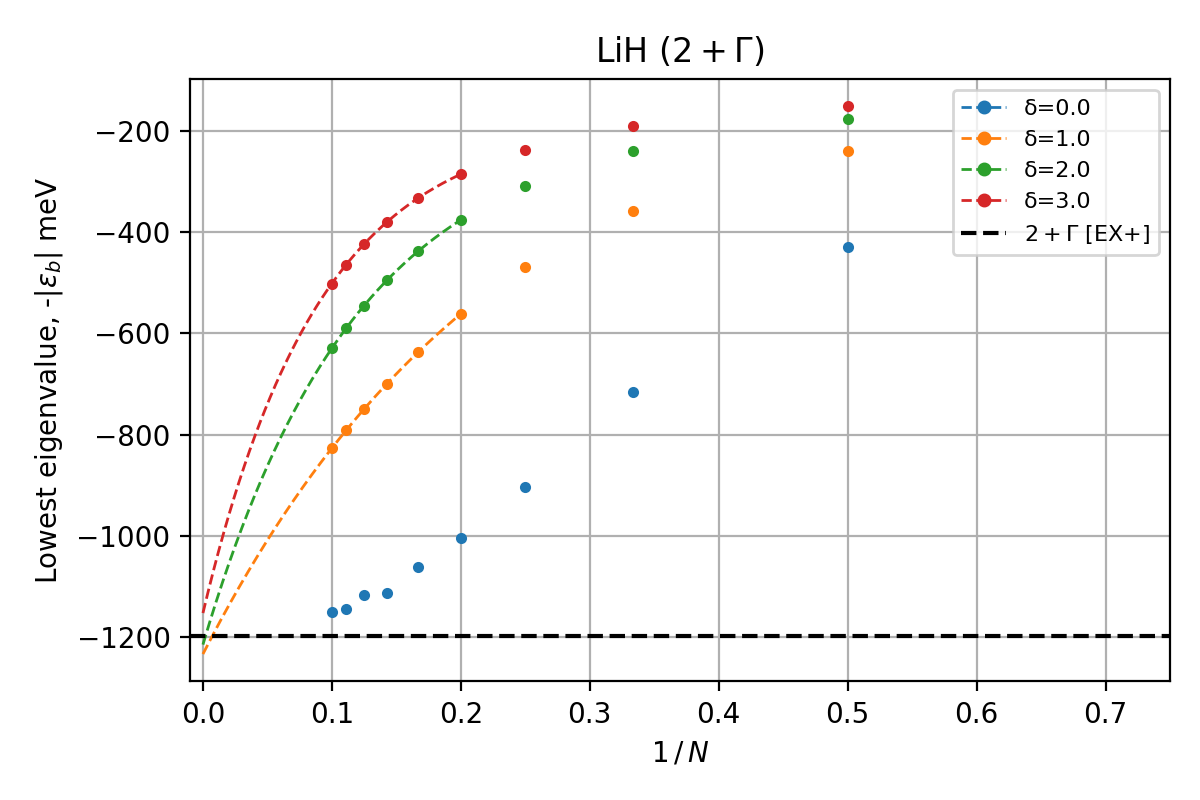}
            \caption{Positron binding energy of LiH calculated via diagMC at the $GW$RPA@TDA, $GW$TDHF@TDA, and $\Sigma^{(2+\Gamma)}$ level of the self energy as a function of the inverse of the maximum diagram order for Ces\`aro--Riesz resummation parameters $\delta = 0$--$3$, as per Eqn.~(9). Dashed red, green and orange curves show exponential extrapolations to $1/N \to 0$. The black horizontal dashed line marks the reference {\tt EXCITON+} Bethe-Salpeter equation solution via exact diagonalisation.\label{fig:1}}
\end{figure}

\begin{figure}[!!th]
    \centering
    \includegraphics[width=0.48\textwidth]{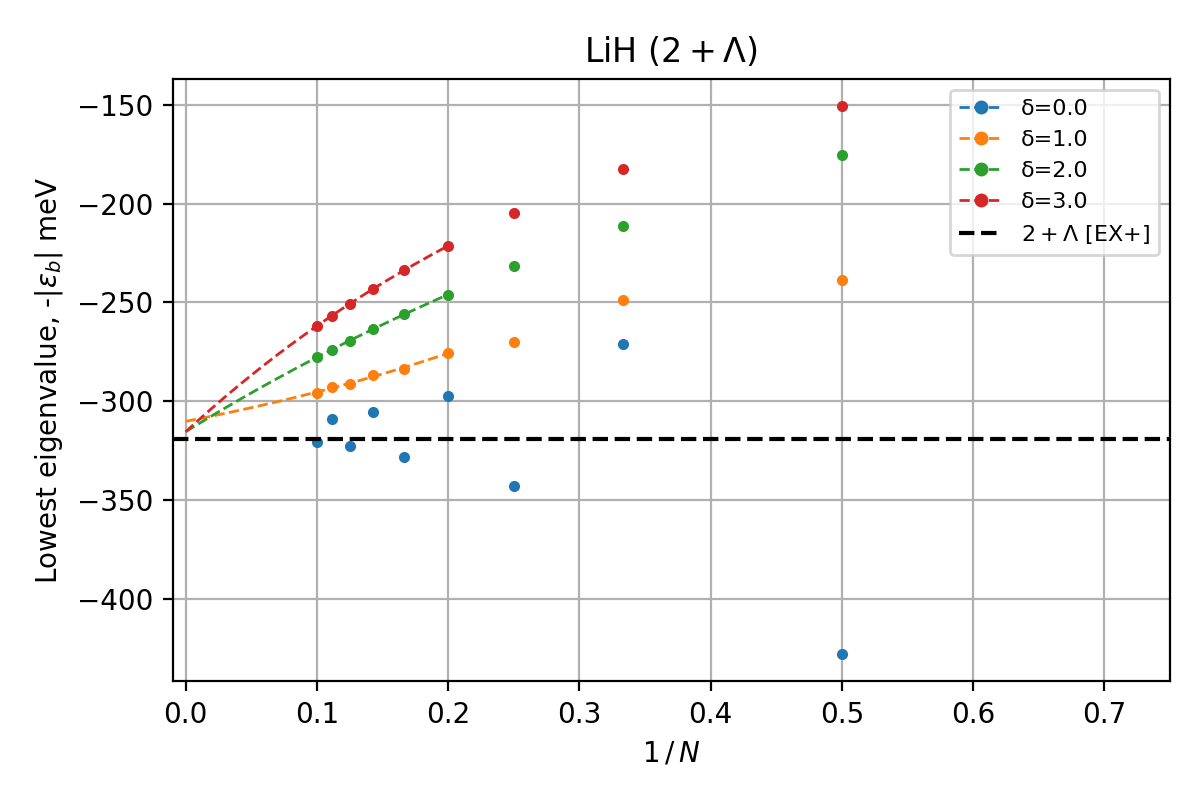}\\
    \includegraphics[width=0.48\textwidth]{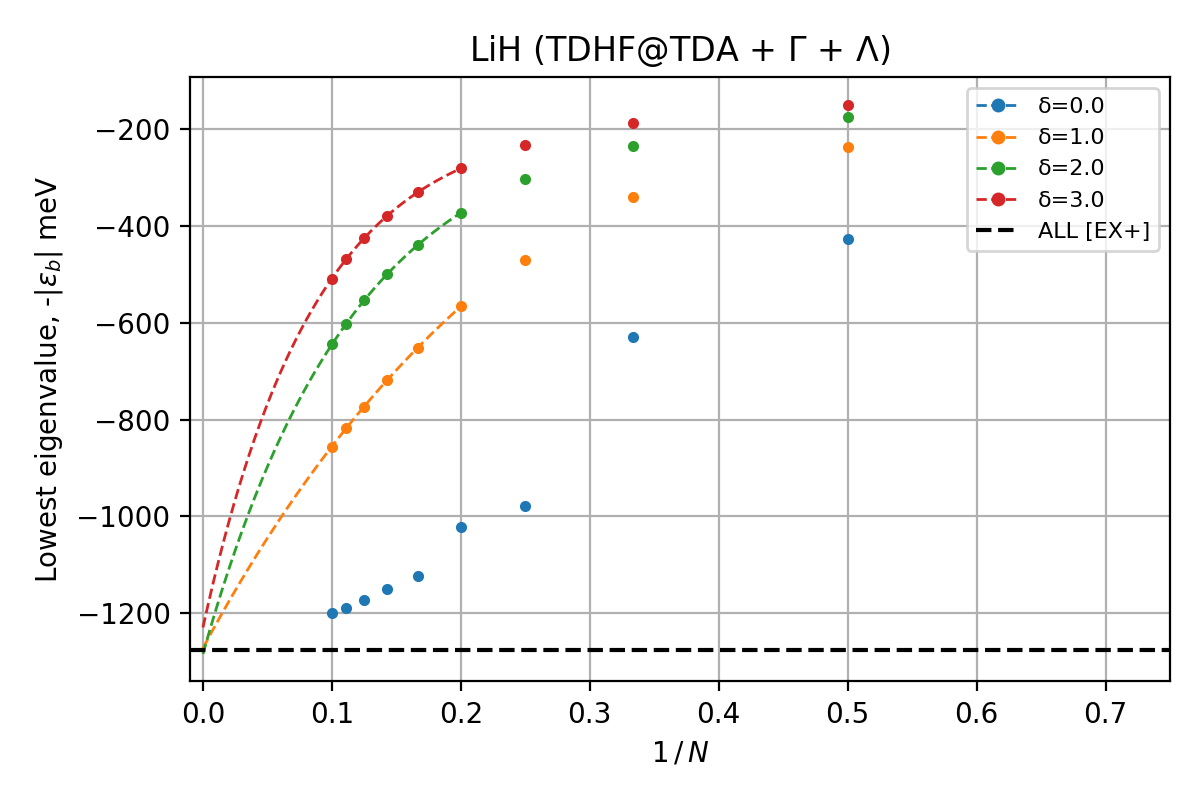}
    \caption{Positron binding energy of LiH calculated via diagMC according to Eqn.~(9) for the positron-hole ladder series ($2+\Lambda$) and at the combined TDHF@TDA${}+\Gamma+\Lambda$ level. Format as in Fig.~\ref{fig:1}. }
    \label{fig:2}
\end{figure}

Figures \ref{fig:1} and \ref{fig:2} show the lowest positron energy eigenvalue (negative of the positron binding $\varepsilon_b>0$) calculated as a function of the inverse of the maximum diagram order for several values of the resummation parameter $\delta$ for the positron-molecule self energy at the RPA@TDA, TDHF@RPA, $\Sigma^{(2+\Gamma)}$ and $\Sigma^{(2+\Lambda)}$ levels. Typically 10$^7$--10$^8$ Monte Carlo steps were used per $\Sigma$ matrix element. 
Table~\ref{tab:binding} summarises the extrapolated binding energies obtained from DiagMC, compared with the values obtained from solution of the Bethe-Salpeter equation via exact diagonalisation in {\tt EXCITON+} . 


For RPA@TDA, TDHF@TDA and the virtual-Ps $\Gamma$-block contribution we find that the series converges monotonically and results in extrapolated binding energies in excellent agreement with the {\tt EXCITON+} benchmarks (see Table \ref{tab:binding}).
We also consider the positron-hole $\Lambda$ ladder series. 
In contrast to the electron-positron ladder series, whose successive rungs contribute equal sign, the positron-hole ladder series is a sign alternating one. Ces\`aro--Riesz resummation produces stable extrapolations to $1/N \to 0$ across the full range $\delta = 0$--$3$. 
Finally, we combine the diagrams to calculate the binding energy in the TDHF@TDA$+\Gamma+\Lambda$ level of theory. The extrapolated binding energy ($1271\pm18$ meV) is again in good agreement with the {\tt EXCITON+} value (1276 meV).

\begin{table}[!!th]
    \centering
    \caption{Extrapolated positron binding energies (meV) from diagMC including the $GW$ random-phase approximation, time-dependent Hatree-Fock approximation within the Tamm-Dancoff approximation, the sum of the second-order diagram and virtual-Ps ladder series ($2+\Gamma$), the second-order diagram and positron-hole ladder series ($2+\Lambda$), and the combined ladders, compared with {\tt EXCITON+} benchmarks.}
    \label{tab:binding}
    \begin{tabular}{ l c c}
        \toprule
        Level & diagMC  & exact diag.  \\
        \midrule
                          $GW$@RPA@TDA  & $376 \pm 0.2$ & $381$  \\
                          $GW$@TDHF@TDA & $636 \pm 1$ & $643$  \\
                          2+$\Gamma$ & $1207 \pm 26$ & $1197$ \\
                          2+$\Lambda$ & $314 \pm 1$ & 319 \\
                          $GW$@TDHF@TDA${}+\Gamma+\Lambda$ & $1271 \pm 18$ & 1276\\
        \bottomrule
    \end{tabular}
\end{table}

The present work provides a proof of principle demonstration of the capability of diagMC to describe the positron-molecule self energy, and specifically the all-order non-perturbative evaluation of the virtual-positronium ladder series, which also contributes to corrections to the annihilation vertex that enhances annihilation rates. 
The approach realises a substantial reduction in memory compared to deterministic calculation via Bethe-Salpeter equations that typically require terabytes of memory, on the order of the size of the orbital basis $N\sim$10$^2$--10$^3$. 
It gives promise for embarrassingly parallel exploitation of modern computing architectures, and for positron binding, scattering and annihilation to be studied in larger molecules than previously feasible,  including via self-consistent construction of diagram series.
The relatively inexpensive exact diagonalisation description of the full $GW$@BSE series (beyond TDA \cite{Hofierka2022}) can be combined with the diagMC evaluation of the virtual-Ps ladder series, or adaptation of sophisticated diagMC approaches e.g., \cite{Chen2019,John2024,Luo2025} to include positrons for atomic and molecular interactions. The approach should be extendable to describe positron localisation and annihilation in condensed matter, relevant to enhancing positron-based diagnostics of industrially-important materials \cite{Tuomisto2013,Hugenschmidt2016} and to the plethora of more general processes involving particle interactions with atoms and molecules, see e.g., \cite{boylepindzola}, by diagMC.

\emph{Data availability.}---All data and the diagMC and development version of the {\tt EXCITON+} codes can be requested from the authors. On acceptance we will make them available freely available fully open-source (under Mozilla license). 

\emph{Author contributions.}---T.A.S. developed the algorithm with D.G.G. T.A.S. developed the code and performed the diagMC calculations, directed by D.G.G. S.K.G. performed the {\tt EXCITON+} exact diagonalisation calculations. D.G.G. additionally conceived the work. All authors contributed to preparing and editing the manuscript. 

\emph{Acknowledgements.}---We thank Brian Cunningham and Gleb Gribakin for helpful discussions and comments on the manuscript. D.G.G thanks Boris Svistunov, Lode Pollet and Evgeny Kozik for encouragement that ultimately led to the current work.  
D.G.G. gratefully acknowledges funding by the European Research Council, grant numbers 804383 and 101170577. T.~A. Scott was funded by a Northern Ireland Department for Economy postgraduate studentship.


\begin{thebibliography}{57}%
\makeatletter
\providecommand \@ifxundefined [1]{%
 \@ifx{#1\undefined}
}%
\providecommand \@ifnum [1]{%
 \ifnum #1\expandafter \@firstoftwo
 \else \expandafter \@secondoftwo
 \fi
}%
\providecommand \@ifx [1]{%
 \ifx #1\expandafter \@firstoftwo
 \else \expandafter \@secondoftwo
 \fi
}%
\providecommand \natexlab [1]{#1}%
\providecommand \enquote  [1]{``#1''}%
\providecommand \bibnamefont  [1]{#1}%
\providecommand \bibfnamefont [1]{#1}%
\providecommand \citenamefont [1]{#1}%
\providecommand \href@noop [0]{\@secondoftwo}%
\providecommand \href [0]{\begingroup \@sanitize@url \@href}%
\providecommand \@href[1]{\@@startlink{#1}\@@href}%
\providecommand \@@href[1]{\endgroup#1\@@endlink}%
\providecommand \@sanitize@url [0]{\catcode `\\12\catcode `\$12\catcode
  `\&12\catcode `\#12\catcode `\^12\catcode `\_12\catcode `\%12\relax}%
\providecommand \@@startlink[1]{}%
\providecommand \@@endlink[0]{}%
\providecommand \url  [0]{\begingroup\@sanitize@url \@url }%
\providecommand \@url [1]{\endgroup\@href {#1}{\urlprefix }}%
\providecommand \urlprefix  [0]{URL }%
\providecommand \Eprint [0]{\href }%
\providecommand \doibase [0]{https://doi.org/}%
\providecommand \selectlanguage [0]{\@gobble}%
\providecommand \bibinfo  [0]{\@secondoftwo}%
\providecommand \bibfield  [0]{\@secondoftwo}%
\providecommand \translation [1]{[#1]}%
\providecommand \BibitemOpen [0]{}%
\providecommand \bibitemStop [0]{}%
\providecommand \bibitemNoStop [0]{.\EOS\space}%
\providecommand \EOS [0]{\spacefactor3000\relax}%
\providecommand \BibitemShut  [1]{\csname bibitem#1\endcsname}%
\let\auto@bib@innerbib\@empty
\bibitem [{\citenamefont {Hofierka}\ \emph {et~al.}(2022)\citenamefont
  {Hofierka}, \citenamefont {Cunningham}, \citenamefont {Rawlins},
  \citenamefont {Patterson},\ and\ \citenamefont {Green}}]{Hofierka2022}%
  \BibitemOpen
  \bibfield  {author} {\bibinfo {author} {\bibfnamefont {J.}~\bibnamefont
  {Hofierka}}, \bibinfo {author} {\bibfnamefont {B.}~\bibnamefont
  {Cunningham}}, \bibinfo {author} {\bibfnamefont {C.~M.}\ \bibnamefont
  {Rawlins}}, \bibinfo {author} {\bibfnamefont {C.~H.}\ \bibnamefont
  {Patterson}},\ and\ \bibinfo {author} {\bibfnamefont {D.~G.}\ \bibnamefont
  {Green}},\ }\bibfield  {title} {\bibinfo {title} {Many-body theory of
  positron binding to polyatomic molecules},\ }\href
  {https://doi.org/10.1038/s41586-022-04703-3} {\bibfield  {journal} {\bibinfo
  {journal} {Nature}\ }\textbf {\bibinfo {volume} {606}},\ \bibinfo {pages}
  {688} (\bibinfo {year} {2022})}\BibitemShut {NoStop}%
\bibitem [{\citenamefont {Tuomisto}\ and\ \citenamefont
  {Makkonen}(2013)}]{Tuomisto2013}%
  \BibitemOpen
  \bibfield  {author} {\bibinfo {author} {\bibfnamefont {F.}~\bibnamefont
  {Tuomisto}}\ and\ \bibinfo {author} {\bibfnamefont {I.}~\bibnamefont
  {Makkonen}},\ }\bibfield  {title} {\bibinfo {title} {Defect identification in
  semiconductors with positron annihilation: Experiment and theory},\
  }\href@noop {} {\bibfield  {journal} {\bibinfo  {journal} {Rev. Mod. Phys.}\
  }\textbf {\bibinfo {volume} {85}},\ \bibinfo {pages} {1583} (\bibinfo {year}
  {2013})}\BibitemShut {NoStop}%
\bibitem [{\citenamefont {Hugenschmidt}(2016)}]{Hugenschmidt2016}%
  \BibitemOpen
  \bibfield  {author} {\bibinfo {author} {\bibfnamefont {C.}~\bibnamefont
  {Hugenschmidt}},\ }\bibfield  {title} {\bibinfo {title} {Positrons in surface
  physics},\ }\href@noop {} {\bibfield  {journal} {\bibinfo  {journal} {Surf.
  Sci. Rep.}\ }\textbf {\bibinfo {volume} {71}},\ \bibinfo {pages} {547}
  (\bibinfo {year} {2016})}\BibitemShut {NoStop}%
\bibitem [{\citenamefont {Danielson}\ \emph {et~al.}(2015)\citenamefont
  {Danielson}, \citenamefont {Dubin}, \citenamefont {Greaves},\ and\
  \citenamefont {Surko}}]{Danielson:2015}%
  \BibitemOpen
  \bibfield  {author} {\bibinfo {author} {\bibfnamefont {J.~R.}\ \bibnamefont
  {Danielson}}, \bibinfo {author} {\bibfnamefont {D.~H.~E.}\ \bibnamefont
  {Dubin}}, \bibinfo {author} {\bibfnamefont {R.~G.}\ \bibnamefont {Greaves}},\
  and\ \bibinfo {author} {\bibfnamefont {C.~M.}\ \bibnamefont {Surko}},\
  }\bibfield  {title} {\bibinfo {title} {Plasma and trap-based techniques for
  science with positrons},\ }\href {https://doi.org/10.1103/RevModPhys.87.247}
  {\bibfield  {journal} {\bibinfo  {journal} {Rev. Mod. Phys.}\ }\textbf
  {\bibinfo {volume} {87}},\ \bibinfo {pages} {247} (\bibinfo {year}
  {2015})}\BibitemShut {NoStop}%
\bibitem [{\citenamefont {Saha}(2005)}]{PETbook2}%
  \BibitemOpen
  \bibfield  {author} {\bibinfo {author} {\bibfnamefont {G.~B.}\ \bibnamefont
  {Saha}},\ }\href@noop {} {\emph {\bibinfo {title} {Basics of PET imaging in
  physics, chemistry, and regulations}}}\ (\bibinfo  {publisher} {Springer, New
  York},\ \bibinfo {year} {2005})\BibitemShut {NoStop}%
\bibitem [{\citenamefont {Wahal}(2008)}]{PETbook}%
  \BibitemOpen
  \bibfield  {author} {\bibinfo {author} {\bibfnamefont {R.~L.}\ \bibnamefont
  {Wahal}},\ }\href@noop {} {\emph {\bibinfo {title} {Principles and Practice
  of Positron Emission Tomography}}}\ (\bibinfo  {publisher} {Lippincott,
  Williams and Wilkins, Philadelphia},\ \bibinfo {year} {2008})\BibitemShut
  {NoStop}%
\bibitem [{\citenamefont {Moskal}\ \emph {et~al.}(2024)\citenamefont {Moskal},
  \citenamefont {Baran} \emph {et~al.}}]{Moskal2024}%
  \BibitemOpen
  \bibfield  {author} {\bibinfo {author} {\bibfnamefont {P.}~\bibnamefont
  {Moskal}}, \bibinfo {author} {\bibfnamefont {J.}~\bibnamefont {Baran}}, \emph
  {et~al.},\ }\bibfield  {title} {\bibinfo {title} {Positronium image of the
  human brain in vivo},\ }\href {https://doi.org/10.1126/sciadv.adp2840}
  {\bibfield  {journal} {\bibinfo  {journal} {Science Advances}\ }\textbf
  {\bibinfo {volume} {10}},\ \bibinfo {pages} {eadp2840} (\bibinfo {year}
  {2024})}\BibitemShut {NoStop}%
\bibitem [{\citenamefont {Moskal}\ \emph {et~al.}(2025)\citenamefont {Moskal},
  \citenamefont {Bilewicz}, \citenamefont {Das}, \citenamefont {Huang},
  \citenamefont {Khreptak}, \citenamefont {Parzych}, \citenamefont {Qi},
  \citenamefont {Rominger}, \citenamefont {Seifert}, \citenamefont {Sharma},
  \citenamefont {Shi}, \citenamefont {Steinberger}, \citenamefont {Walczak},\
  and\ \citenamefont {{St\ifmmode \mbox{\k{e}}\else \k{e}\fi{}pie\ifmmode
  \acute{n}\else \'{n}\fi{}}}}]{JPET2025}%
  \BibitemOpen
  \bibfield  {author} {\bibinfo {author} {\bibfnamefont {P.}~\bibnamefont
  {Moskal}}, \bibinfo {author} {\bibfnamefont {A.}~\bibnamefont {Bilewicz}},
  \bibinfo {author} {\bibfnamefont {M.}~\bibnamefont {Das}}, \bibinfo {author}
  {\bibfnamefont {B.}~\bibnamefont {Huang}}, \bibinfo {author} {\bibfnamefont
  {A.}~\bibnamefont {Khreptak}}, \bibinfo {author} {\bibfnamefont
  {S.}~\bibnamefont {Parzych}}, \bibinfo {author} {\bibfnamefont
  {J.}~\bibnamefont {Qi}}, \bibinfo {author} {\bibfnamefont {A.}~\bibnamefont
  {Rominger}}, \bibinfo {author} {\bibfnamefont {R.}~\bibnamefont {Seifert}},
  \bibinfo {author} {\bibfnamefont {S.}~\bibnamefont {Sharma}}, \bibinfo
  {author} {\bibfnamefont {K.}~\bibnamefont {Shi}}, \bibinfo {author}
  {\bibfnamefont {W.~M.}\ \bibnamefont {Steinberger}}, \bibinfo {author}
  {\bibfnamefont {R.}~\bibnamefont {Walczak}},\ and\ \bibinfo {author}
  {\bibfnamefont {E.}~\bibnamefont {{St\ifmmode \mbox{\k{e}}\else
  \k{e}\fi{}pie\ifmmode \acute{n}\else \'{n}\fi{}}}},\ }\bibfield  {title}
  {\bibinfo {title} {Positronium imaging: History, current status, and future
  perspectives},\ }\href {https://doi.org/10.1109/TRPMS.2025.3583554}
  {\bibfield  {journal} {\bibinfo  {journal} {IEEE Transactions on Radiation
  and Plasma Medical Sciences}\ }\textbf {\bibinfo {volume} {9}},\ \bibinfo
  {pages} {981} (\bibinfo {year} {2025})}\BibitemShut {NoStop}%
\bibitem [{\citenamefont {Drachman}(1996)}]{posfun}%
  \BibitemOpen
  \bibfield  {author} {\bibinfo {author} {\bibfnamefont {R.~J.}\ \bibnamefont
  {Drachman}},\ }\bibfield  {title} {\bibinfo {title} {{Why positron physics is
  fun}},\ }\href {https://doi.org/10.1063/1.49828} {\bibfield  {journal}
  {\bibinfo  {journal} {AIP Conference Proceedings}\ }\textbf {\bibinfo
  {volume} {360}},\ \bibinfo {pages} {369} (\bibinfo {year}
  {1996})}\BibitemShut {NoStop}%
\bibitem [{\citenamefont {Prantzos}\ \emph {et~al.}(2011)\citenamefont
  {Prantzos}, \citenamefont {Boehm}, \citenamefont {Bykov}, \citenamefont
  {Diehl}, \citenamefont {Ferrière}, \citenamefont {Guessoum}, \citenamefont
  {Jean}, \citenamefont {Knoedlseder}, \citenamefont {Marcowith}, \citenamefont
  {Moskalenko}, \citenamefont {Strong},\ and\ \citenamefont
  {Weidenspointner}}]{prantzos_511_2011}%
  \BibitemOpen
  \bibfield  {author} {\bibinfo {author} {\bibfnamefont {N.}~\bibnamefont
  {Prantzos}}, \bibinfo {author} {\bibfnamefont {C.}~\bibnamefont {Boehm}},
  \bibinfo {author} {\bibfnamefont {A.~M.}\ \bibnamefont {Bykov}}, \bibinfo
  {author} {\bibfnamefont {R.}~\bibnamefont {Diehl}}, \bibinfo {author}
  {\bibfnamefont {K.}~\bibnamefont {Ferrière}}, \bibinfo {author}
  {\bibfnamefont {N.}~\bibnamefont {Guessoum}}, \bibinfo {author}
  {\bibfnamefont {P.}~\bibnamefont {Jean}}, \bibinfo {author} {\bibfnamefont
  {J.}~\bibnamefont {Knoedlseder}}, \bibinfo {author} {\bibfnamefont
  {A.}~\bibnamefont {Marcowith}}, \bibinfo {author} {\bibfnamefont {I.~V.}\
  \bibnamefont {Moskalenko}}, \bibinfo {author} {\bibfnamefont
  {A.}~\bibnamefont {Strong}},\ and\ \bibinfo {author} {\bibfnamefont
  {G.}~\bibnamefont {Weidenspointner}},\ }\bibfield  {title} {\bibinfo {title}
  {The 511 {keV} emission from positron annihilation in the {Galaxy}},\ }\href
  {https://doi.org/10.1103/RevModPhys.83.1001} {\bibfield  {journal} {\bibinfo
  {journal} {Rev. Mod. Phys.}\ }\textbf {\bibinfo {volume} {83}},\ \bibinfo
  {pages} {1001} (\bibinfo {year} {2011})},\ \bibinfo {note} {publisher:
  American Physical Society}\BibitemShut {NoStop}%
\bibitem [{\citenamefont {Fuller}\ \emph {et~al.}(2019)\citenamefont {Fuller},
  \citenamefont {Kusenko}, \citenamefont {Radice},\ and\ \citenamefont
  {Takhistov}}]{Fuller:2019}%
  \BibitemOpen
  \bibfield  {author} {\bibinfo {author} {\bibfnamefont {G.~M.}\ \bibnamefont
  {Fuller}}, \bibinfo {author} {\bibfnamefont {A.}~\bibnamefont {Kusenko}},
  \bibinfo {author} {\bibfnamefont {D.}~\bibnamefont {Radice}},\ and\ \bibinfo
  {author} {\bibfnamefont {V.}~\bibnamefont {Takhistov}},\ }\bibfield  {title}
  {\bibinfo {title} {Positrons and 511 kev radiation as tracers of recent
  binary neutron star mergers},\ }\href
  {https://doi.org/10.1103/PhysRevLett.122.121101} {\bibfield  {journal}
  {\bibinfo  {journal} {Phys. Rev. Lett.}\ }\textbf {\bibinfo {volume} {122}},\
  \bibinfo {pages} {121101} (\bibinfo {year} {2019})}\BibitemShut {NoStop}%
\bibitem [{\citenamefont {Flambaum}\ and\ \citenamefont
  {Samsonov}(2021)}]{Flambaum:2021}%
  \BibitemOpen
  \bibfield  {author} {\bibinfo {author} {\bibfnamefont {V.~V.}\ \bibnamefont
  {Flambaum}}\ and\ \bibinfo {author} {\bibfnamefont {I.~B.}\ \bibnamefont
  {Samsonov}},\ }\bibfield  {title} {\bibinfo {title} {Radiation from
  matter-antimatter annihilation in the quark nugget model of dark matter},\
  }\href {https://doi.org/10.1103/PhysRevD.104.063042} {\bibfield  {journal}
  {\bibinfo  {journal} {Phys. Rev. D}\ }\textbf {\bibinfo {volume} {104}},\
  \bibinfo {pages} {063042} (\bibinfo {year} {2021})}\BibitemShut {NoStop}%
\bibitem [{\citenamefont {Gribakin}\ \emph {et~al.}(2010)\citenamefont
  {Gribakin}, \citenamefont {Young},\ and\ \citenamefont
  {Surko}}]{RevModPhys.82.2557}%
  \BibitemOpen
  \bibfield  {author} {\bibinfo {author} {\bibfnamefont {G.~F.}\ \bibnamefont
  {Gribakin}}, \bibinfo {author} {\bibfnamefont {J.~A.}\ \bibnamefont
  {Young}},\ and\ \bibinfo {author} {\bibfnamefont {C.~M.}\ \bibnamefont
  {Surko}},\ }\bibfield  {title} {\bibinfo {title} {Positron-molecule
  interactions: Resonant attachment, annihilation, and bound states},\ }\href
  {https://doi.org/10.1103/RevModPhys.82.2557} {\bibfield  {journal} {\bibinfo
  {journal} {Rev. Mod. Phys.}\ }\textbf {\bibinfo {volume} {82}},\ \bibinfo
  {pages} {2557} (\bibinfo {year} {2010})}\BibitemShut {NoStop}%
\bibitem [{\citenamefont {Gribakin}\ \emph {et~al.}(2017)\citenamefont
  {Gribakin}, \citenamefont {Stanton}, \citenamefont {Danielson}, \citenamefont
  {Natisin},\ and\ \citenamefont {Surko}}]{Gribakin2017}%
  \BibitemOpen
  \bibfield  {author} {\bibinfo {author} {\bibfnamefont {G.~F.}\ \bibnamefont
  {Gribakin}}, \bibinfo {author} {\bibfnamefont {J.~F.}\ \bibnamefont
  {Stanton}}, \bibinfo {author} {\bibfnamefont {J.~R.}\ \bibnamefont
  {Danielson}}, \bibinfo {author} {\bibfnamefont {M.~R.}\ \bibnamefont
  {Natisin}},\ and\ \bibinfo {author} {\bibfnamefont {C.~M.}\ \bibnamefont
  {Surko}},\ }\bibfield  {title} {\bibinfo {title} {Mode coupling and
  multiquantum vibrational excitations in feshbach-resonant positron
  annihilation in molecules},\ }\href
  {https://doi.org/10.1103/PhysRevA.96.062709} {\bibfield  {journal} {\bibinfo
  {journal} {Phys. Rev. A}\ }\textbf {\bibinfo {volume} {96}},\ \bibinfo
  {pages} {062709} (\bibinfo {year} {2017})}\BibitemShut {NoStop}%
\bibitem [{\citenamefont {Boyle}\ and\ \citenamefont
  {Pindzola}(1998)}]{boylepindzola}%
  \BibitemOpen
  \bibfield  {author} {\bibinfo {author} {\bibfnamefont {J.}~\bibnamefont
  {Boyle}}\ and\ \bibinfo {author} {\bibfnamefont {M.}~\bibnamefont
  {Pindzola}},\ }\href@noop {} {\emph {\bibinfo {title} {Many-body atomic
  physics}}}\ (\bibinfo  {publisher} {Cambridge University Press},\ \bibinfo
  {year} {1998})\BibitemShut {NoStop}%
\bibitem [{\citenamefont {Amusia}\ \emph {et~al.}(1976)\citenamefont {Amusia},
  \citenamefont {Cherepkov}, \citenamefont {Chernysheva},\ and\ \citenamefont
  {Shapiro}}]{Amusia:Pos:MBT:He}%
  \BibitemOpen
  \bibfield  {author} {\bibinfo {author} {\bibfnamefont {M.~Y.}\ \bibnamefont
  {Amusia}}, \bibinfo {author} {\bibfnamefont {N.~A.}\ \bibnamefont
  {Cherepkov}}, \bibinfo {author} {\bibfnamefont {L.~V.}\ \bibnamefont
  {Chernysheva}},\ and\ \bibinfo {author} {\bibfnamefont {S.~G.}\ \bibnamefont
  {Shapiro}},\ }\bibfield  {title} {\bibinfo {title} {Elastic scattering of
  slow positrons by helium},\ }\href
  {http://stacks.iop.org/0022-3700/9/i=17/a=005} {\bibfield  {journal}
  {\bibinfo  {journal} {J. Phys. B: Atom. Mol. Phys.}\ }\textbf {\bibinfo
  {volume} {9}},\ \bibinfo {pages} {L531} (\bibinfo {year} {1976})}\BibitemShut
  {NoStop}%
\bibitem [{\citenamefont {Dzuba}\ \emph {et~al.}(1993)\citenamefont {Dzuba},
  \citenamefont {Flambaum}, \citenamefont {King}, \citenamefont {Miller},\ and\
  \citenamefont {Sushkov}}]{PhysScripta.46.248}%
  \BibitemOpen
  \bibfield  {author} {\bibinfo {author} {\bibfnamefont {V.~A.}\ \bibnamefont
  {Dzuba}}, \bibinfo {author} {\bibfnamefont {V.~V.}\ \bibnamefont {Flambaum}},
  \bibinfo {author} {\bibfnamefont {W.~A.}\ \bibnamefont {King}}, \bibinfo
  {author} {\bibfnamefont {B.~N.}\ \bibnamefont {Miller}},\ and\ \bibinfo
  {author} {\bibfnamefont {O.~P.}\ \bibnamefont {Sushkov}},\ }\bibfield
  {title} {\bibinfo {title} {Interaction between slow positrons and atoms},\
  }\href {https://doi.org/10.1088/0031-8949/1993/T46/039} {\bibfield  {journal}
  {\bibinfo  {journal} {Phys. Scr.}\ }\textbf {\bibinfo {volume} {T46}},\
  \bibinfo {pages} {248} (\bibinfo {year} {1993})}\BibitemShut {NoStop}%
\bibitem [{\citenamefont {Dzuba}\ \emph {et~al.}(1995)\citenamefont {Dzuba},
  \citenamefont {Flambaum}, \citenamefont {Gribakin},\ and\ \citenamefont
  {King}}]{PhysRevA.52.4541}%
  \BibitemOpen
  \bibfield  {author} {\bibinfo {author} {\bibfnamefont {V.~A.}\ \bibnamefont
  {Dzuba}}, \bibinfo {author} {\bibfnamefont {V.~V.}\ \bibnamefont {Flambaum}},
  \bibinfo {author} {\bibfnamefont {G.~F.}\ \bibnamefont {Gribakin}},\ and\
  \bibinfo {author} {\bibfnamefont {W.~A.}\ \bibnamefont {King}},\ }\bibfield
  {title} {\bibinfo {title} {Bound states of positrons and neutral atoms},\
  }\href {https://doi.org/10.1103/PhysRevA.52.4541} {\bibfield  {journal}
  {\bibinfo  {journal} {Phys. Rev. A}\ }\textbf {\bibinfo {volume} {52}},\
  \bibinfo {pages} {4541} (\bibinfo {year} {1995})}\BibitemShut {NoStop}%
\bibitem [{\citenamefont {Gribakin}\ and\ \citenamefont
  {Ludlow}(2004)}]{Gribakin:2004}%
  \BibitemOpen
  \bibfield  {author} {\bibinfo {author} {\bibfnamefont {G.~F.}\ \bibnamefont
  {Gribakin}}\ and\ \bibinfo {author} {\bibfnamefont {J.}~\bibnamefont
  {Ludlow}},\ }\bibfield  {title} {\bibinfo {title} {Many-body theory of
  positron-atom interactions},\ }\href
  {https://doi.org/10.1103/PhysRevA.70.032720} {\bibfield  {journal} {\bibinfo
  {journal} {Phys. Rev. A}\ }\textbf {\bibinfo {volume} {70}},\ \bibinfo
  {pages} {032720} (\bibinfo {year} {2004})}\BibitemShut {NoStop}%
\bibitem [{\citenamefont {Harabati}\ \emph {et~al.}(2014)\citenamefont
  {Harabati}, \citenamefont {Dzuba},\ and\ \citenamefont
  {Flambaum}}]{harabati2014identification}%
  \BibitemOpen
  \bibfield  {author} {\bibinfo {author} {\bibfnamefont {C.}~\bibnamefont
  {Harabati}}, \bibinfo {author} {\bibfnamefont {V.}~\bibnamefont {Dzuba}},\
  and\ \bibinfo {author} {\bibfnamefont {V.}~\bibnamefont {Flambaum}},\
  }\bibfield  {title} {\bibinfo {title} {Identification of atoms that can bind
  positrons},\ }\href@noop {} {\bibfield  {journal} {\bibinfo  {journal} {Phys.
  Rev. A.}\ }\textbf {\bibinfo {volume} {89}},\ \bibinfo {pages} {022517}
  (\bibinfo {year} {2014})}\BibitemShut {NoStop}%
\bibitem [{\citenamefont {M\"uller}\ and\ \citenamefont
  {Cederbaum}(1990)}]{Cederbaum-elecpos}%
  \BibitemOpen
  \bibfield  {author} {\bibinfo {author} {\bibfnamefont {M.}~\bibnamefont
  {M\"uller}}\ and\ \bibinfo {author} {\bibfnamefont {L.~S.}\ \bibnamefont
  {Cederbaum}},\ }\bibfield  {title} {\bibinfo {title} {Many-body theory of
  composite electronic-positronic systems},\ }\href
  {https://doi.org/10.1103/PhysRevA.42.170} {\bibfield  {journal} {\bibinfo
  {journal} {Phys. Rev. A}\ }\textbf {\bibinfo {volume} {42}},\ \bibinfo
  {pages} {170} (\bibinfo {year} {1990})}\BibitemShut {NoStop}%
\bibitem [{\citenamefont {Green}\ and\ \citenamefont
  {Gribakin}(2013)}]{Green2013}%
  \BibitemOpen
  \bibfield  {author} {\bibinfo {author} {\bibfnamefont {D.~G.}\ \bibnamefont
  {Green}}\ and\ \bibinfo {author} {\bibfnamefont {G.~F.}\ \bibnamefont
  {Gribakin}},\ }\bibfield  {title} {\bibinfo {title} {Positron scattering and
  annihilation in hydrogenlike ions},\ }\href
  {https://doi.org/10.1103/PhysRevA.88.032708} {\bibfield  {journal} {\bibinfo
  {journal} {Phys. Rev. A}\ }\textbf {\bibinfo {volume} {88}},\ \bibinfo
  {pages} {032708} (\bibinfo {year} {2013})}\BibitemShut {NoStop}%
\bibitem [{\citenamefont {Green}\ \emph {et~al.}(2014)\citenamefont {Green},
  \citenamefont {Ludlow},\ and\ \citenamefont {Gribakin}}]{Green:2014}%
  \BibitemOpen
  \bibfield  {author} {\bibinfo {author} {\bibfnamefont {D.~G.}\ \bibnamefont
  {Green}}, \bibinfo {author} {\bibfnamefont {J.~A.}\ \bibnamefont {Ludlow}},\
  and\ \bibinfo {author} {\bibfnamefont {G.~F.}\ \bibnamefont {Gribakin}},\
  }\bibfield  {title} {\bibinfo {title} {Positron scattering and annihilation
  on noble-gas atoms},\ }\href {https://doi.org/10.1103/PhysRevA.90.032712}
  {\bibfield  {journal} {\bibinfo  {journal} {Phys. Rev. A}\ }\textbf {\bibinfo
  {volume} {90}},\ \bibinfo {pages} {032712} (\bibinfo {year}
  {2014})}\BibitemShut {NoStop}%
\bibitem [{\citenamefont {Green}\ and\ \citenamefont
  {Gribakin}(2015)}]{Green2015}%
  \BibitemOpen
  \bibfield  {author} {\bibinfo {author} {\bibfnamefont {D.~G.}\ \bibnamefont
  {Green}}\ and\ \bibinfo {author} {\bibfnamefont {G.~F.}\ \bibnamefont
  {Gribakin}},\ }\bibfield  {title} {\bibinfo {title} {$\ensuremath{\gamma}$
  spectra and enhancement factors for positron annihilation with core
  electrons},\ }\href {https://doi.org/10.1103/PhysRevLett.114.093201}
  {\bibfield  {journal} {\bibinfo  {journal} {Phys. Rev. Lett.}\ }\textbf
  {\bibinfo {volume} {114}},\ \bibinfo {pages} {093201} (\bibinfo {year}
  {2015})}\BibitemShut {NoStop}%
\bibitem [{\citenamefont {Rawlins}\ \emph {et~al.}(2023)\citenamefont
  {Rawlins}, \citenamefont {Hofierka}, \citenamefont {Cunningham},
  \citenamefont {Patterson},\ and\ \citenamefont {Green}}]{Rawlins2023}%
  \BibitemOpen
  \bibfield  {author} {\bibinfo {author} {\bibfnamefont {C.~M.}\ \bibnamefont
  {Rawlins}}, \bibinfo {author} {\bibfnamefont {J.}~\bibnamefont {Hofierka}},
  \bibinfo {author} {\bibfnamefont {B.}~\bibnamefont {Cunningham}}, \bibinfo
  {author} {\bibfnamefont {C.~H.}\ \bibnamefont {Patterson}},\ and\ \bibinfo
  {author} {\bibfnamefont {D.~G.}\ \bibnamefont {Green}},\ }\bibfield  {title}
  {\bibinfo {title} {Many-body theory calculations of positron scattering and
  annihilation in {${\mathrm{H}}_{2}$}, {${\mathrm{N}}_{2}$}, and
  {${\mathrm{CH}}_{4}$}},\ }\href
  {https://doi.org/10.1103/PhysRevLett.130.263001} {\bibfield  {journal}
  {\bibinfo  {journal} {Phys. Rev. Lett.}\ }\textbf {\bibinfo {volume} {130}},\
  \bibinfo {pages} {263001} (\bibinfo {year} {2023})}\BibitemShut {NoStop}%
\bibitem [{\citenamefont {Fetter}\ and\ \citenamefont
  {Walecka}(1971)}]{Fetter}%
  \BibitemOpen
  \bibfield  {author} {\bibinfo {author} {\bibfnamefont {A.~L.}\ \bibnamefont
  {Fetter}}\ and\ \bibinfo {author} {\bibfnamefont {J.~D.}\ \bibnamefont
  {Walecka}},\ }\href@noop {} {\emph {\bibinfo {title} {Quantum Theory of
  Many-Particle Systems}}}\ (\bibinfo  {publisher} {McGraw-Hill},\ \bibinfo
  {year} {1971})\BibitemShut {NoStop}%
\bibitem [{\citenamefont {Dickhoff}\ and\ \citenamefont
  {Van~Neck}(2025)}]{mbtexposed}%
  \BibitemOpen
  \bibfield  {author} {\bibinfo {author} {\bibfnamefont {W.~H.}\ \bibnamefont
  {Dickhoff}}\ and\ \bibinfo {author} {\bibfnamefont {D.}~\bibnamefont
  {Van~Neck}},\ }\href {https://doi.org/10.1142/14178} {\emph {\bibinfo {title}
  {Many-Body Theory Exposed!}}},\ \bibinfo {edition} {3rd}\ ed.\ (\bibinfo
  {publisher} {World Scientific},\ \bibinfo {year} {2025})\BibitemShut
  {NoStop}%
\bibitem [{Note1()}]{Note1}%
  \BibitemOpen
  \bibinfo {note} {For example, for a virtual-Ps ladder contribution to the
  positron-molecule self energy with $n$ intermediate electron-positron interactions, i.e., at $(n+2)^{th}$ order in the electron-positron
  Coulomb interaction with spectator hole, one must sum over $N^{2(n+1)}$
  combinations of the number of excited electron and positron intermediate states
  (MOs) in products of Coulomb matrix elements and energy denominators $\DOTSB \sum@ \slimits@ _{\mu _1\nu _1\protect \dots \mu
  _{n+1}\nu _{n+1}} \Pi _{k=1}^n \protect V^k_{\mu _k\nu
  _k,\mu _{k+1}\nu _{k+1}} / \Delta ^{k+1}_E$. Typically $N\gtrsim
  10^2$, thus the number of combinations grows beyond that capable for deterministic calculation swiftly}\BibitemShut {NoStop}%
\bibitem [{Note2()}]{Note2}%
  \BibitemOpen
  \bibinfo {note} {For the positron-molecule problem the $GW$ diagram alone is
  wholly deficient. The importance of the virtual-positronium $\Gamma $ ladder
  series arises from the fact that successive terms in the series contribute
  with equal sign, in contrast to the all-electron case in which the signs
  alternate, leading to substantial cancellation. The virtual-positronium
  ladder series also corrects the annihilation amplitude vertex and its
  evaluation is crucial for accurate calculations of positron and positronium
  annihilation rates \cite {Green2015,Green2018,Swann2023}.}\BibitemShut
  {Stop}%
\bibitem [{\citenamefont {Cassidy}\ \emph
  {et~al.}(2024{\natexlab{a}})\citenamefont {Cassidy}, \citenamefont
  {Hofierka}, \citenamefont {Cunningham}, \citenamefont {Rawlins},
  \citenamefont {Patterson},\ and\ \citenamefont {Green}}]{Cassidy2024}%
  \BibitemOpen
  \bibfield  {author} {\bibinfo {author} {\bibfnamefont {J.~P.}\ \bibnamefont
  {Cassidy}}, \bibinfo {author} {\bibfnamefont {J.}~\bibnamefont {Hofierka}},
  \bibinfo {author} {\bibfnamefont {B.}~\bibnamefont {Cunningham}}, \bibinfo
  {author} {\bibfnamefont {C.~M.}\ \bibnamefont {Rawlins}}, \bibinfo {author}
  {\bibfnamefont {C.~H.}\ \bibnamefont {Patterson}},\ and\ \bibinfo {author}
  {\bibfnamefont {D.~G.}\ \bibnamefont {Green}},\ }\bibfield  {title} {\bibinfo
  {title} {Many-body theory calculations of positron binding to halogenated
  hydrocarbons},\ }\href {https://doi.org/10.1103/PhysRevA.109.L040801}
  {\bibfield  {journal} {\bibinfo  {journal} {Phys. Rev. A}\ }\textbf {\bibinfo
  {volume} {109}},\ \bibinfo {pages} {L040801} (\bibinfo {year}
  {2024}{\natexlab{a}})}\BibitemShut {NoStop}%
\bibitem [{\citenamefont {Hofierka}\ \emph {et~al.}(2024)\citenamefont
  {Hofierka}, \citenamefont {Cunningham},\ and\ \citenamefont
  {Green}}]{Hofierka2024}%
  \BibitemOpen
  \bibfield  {author} {\bibinfo {author} {\bibfnamefont {J.}~\bibnamefont
  {Hofierka}}, \bibinfo {author} {\bibfnamefont {B.}~\bibnamefont
  {Cunningham}},\ and\ \bibinfo {author} {\bibfnamefont {D.~G.}\ \bibnamefont
  {Green}},\ }\bibfield  {title} {\bibinfo {title} {Many-body theory
  calculations of positron binding to hydrogen cyanide},\ }\href@noop {}
  {\bibfield  {journal} {\bibinfo  {journal} {Eur. Phys. J. D}\ }\textbf
  {\bibinfo {volume} {78}},\ \bibinfo {pages} {37} (\bibinfo {year}
  {2024})}\BibitemShut {NoStop}%
\bibitem [{\citenamefont {Arthur-Baidoo}\ \emph {et~al.}(2024)\citenamefont
  {Arthur-Baidoo}, \citenamefont {Danielson}, \citenamefont {Surko},
  \citenamefont {Cassidy}, \citenamefont {Gregg}, \citenamefont {Hofierka},
  \citenamefont {Cunningham}, \citenamefont {Patterson},\ and\ \citenamefont
  {Green}}]{ArthurBaidoo2024}%
  \BibitemOpen
  \bibfield  {author} {\bibinfo {author} {\bibfnamefont {E.}~\bibnamefont
  {Arthur-Baidoo}}, \bibinfo {author} {\bibfnamefont {J.~R.}\ \bibnamefont
  {Danielson}}, \bibinfo {author} {\bibfnamefont {C.~M.}\ \bibnamefont
  {Surko}}, \bibinfo {author} {\bibfnamefont {J.~P.}\ \bibnamefont {Cassidy}},
  \bibinfo {author} {\bibfnamefont {S.~K.}\ \bibnamefont {Gregg}}, \bibinfo
  {author} {\bibfnamefont {J.}~\bibnamefont {Hofierka}}, \bibinfo {author}
  {\bibfnamefont {B.}~\bibnamefont {Cunningham}}, \bibinfo {author}
  {\bibfnamefont {C.~H.}\ \bibnamefont {Patterson}},\ and\ \bibinfo {author}
  {\bibfnamefont {D.~G.}\ \bibnamefont {Green}},\ }\bibfield  {title} {\bibinfo
  {title} {Positron annihilation and binding in aromatic and other ring
  molecules},\ }\href {https://doi.org/10.1103/PhysRevA.109.062801} {\bibfield
  {journal} {\bibinfo  {journal} {Phys. Rev. A}\ }\textbf {\bibinfo {volume}
  {109}},\ \bibinfo {pages} {062801} (\bibinfo {year} {2024})}\BibitemShut
  {NoStop}%
\bibitem [{\citenamefont {Gregg}\ \emph
  {et~al.}(2025{\natexlab{a}})\citenamefont {Gregg}, \citenamefont {Hofierka},
  \citenamefont {Cunningham},\ and\ \citenamefont {Green}}]{Gregg2025}%
  \BibitemOpen
  \bibfield  {author} {\bibinfo {author} {\bibfnamefont {S.~K.}\ \bibnamefont
  {Gregg}}, \bibinfo {author} {\bibfnamefont {J.}~\bibnamefont {Hofierka}},
  \bibinfo {author} {\bibfnamefont {B.}~\bibnamefont {Cunningham}},\ and\
  \bibinfo {author} {\bibfnamefont {D.~G.}\ \bibnamefont {Green}},\ }\href@noop
  {} {\bibinfo {title} {Many-body theory calculations of positron binding to
  parabenzoquinone}} (\bibinfo {year} {2025}{\natexlab{a}}),\ \Eprint
  {https://arxiv.org/abs/2502.10327} {arXiv:2502.10327} \BibitemShut {NoStop}%
\bibitem [{\citenamefont {Gregg}\ and\ \citenamefont
  {Green}(2026)}]{Gregg2026}%
  \BibitemOpen
  \bibfield  {author} {\bibinfo {author} {\bibfnamefont {S.~K.}\ \bibnamefont
  {Gregg}}\ and\ \bibinfo {author} {\bibfnamefont {D.~G.}\ \bibnamefont
  {Green}},\ }\bibfield  {title} {\bibinfo {title} {Many-body theory
  predictions of positron binding energies in five-membered heterocycles
  involving {N}, {O}, {S}, and {NH} substituents},\ }\bibfield  {journal}
  {\bibinfo  {journal} {J. Chem. Theory Comput.}\ }\href
  {https://doi.org/10.1021/acs.jctc.6c00759} {10.1021/acs.jctc.6c00759}
  (\bibinfo {year} {2026})\BibitemShut {NoStop}%
\bibitem [{\citenamefont {Hofierka}\ \emph {et~al.}(2023)\citenamefont
  {Hofierka}, \citenamefont {Rawlins}, \citenamefont {Cunningham},
  \citenamefont {Waide},\ and\ \citenamefont {Green}}]{Hofierka2023}%
  \BibitemOpen
  \bibfield  {author} {\bibinfo {author} {\bibfnamefont {J.}~\bibnamefont
  {Hofierka}}, \bibinfo {author} {\bibfnamefont {C.~M.}\ \bibnamefont
  {Rawlins}}, \bibinfo {author} {\bibfnamefont {B.}~\bibnamefont {Cunningham}},
  \bibinfo {author} {\bibfnamefont {D.~T.}\ \bibnamefont {Waide}},\ and\
  \bibinfo {author} {\bibfnamefont {D.~G.}\ \bibnamefont {Green}},\ }\bibfield
  {title} {\bibinfo {title} {Many-body theory calculations of positron
  scattering and annihilation in noble-gas atoms via the solution of
  {B}ethe–{S}alpeter equations using the gaussian-basis code {EXCITON+}},\
  }\href@noop {} {\bibfield  {journal} {\bibinfo  {journal} {Front.~in
  Physics}\ }\textbf {\bibinfo {volume} {11}} (\bibinfo {year}
  {2023})}\BibitemShut {NoStop}%
\bibitem [{\citenamefont {Gregg}\ \emph
  {et~al.}(2025{\natexlab{b}})\citenamefont {Gregg}, \citenamefont {Cassidy},
  \citenamefont {Swann}, \citenamefont {Hofierka}, \citenamefont {Cunningham},\
  and\ \citenamefont {Green}}]{Gregg2025Gamma}%
  \BibitemOpen
  \bibfield  {author} {\bibinfo {author} {\bibfnamefont {S.~K.}\ \bibnamefont
  {Gregg}}, \bibinfo {author} {\bibfnamefont {J.~P.}\ \bibnamefont {Cassidy}},
  \bibinfo {author} {\bibfnamefont {A.~R.}\ \bibnamefont {Swann}}, \bibinfo
  {author} {\bibfnamefont {J.}~\bibnamefont {Hofierka}}, \bibinfo {author}
  {\bibfnamefont {B.}~\bibnamefont {Cunningham}},\ and\ \bibinfo {author}
  {\bibfnamefont {D.~G.}\ \bibnamefont {Green}},\ }\href@noop {} {\bibinfo
  {title} {Many-body theory and gaussian-basis implementation of positron
  annihilation $\gamma$-ray spectra on polyatomic molecules}} (\bibinfo {year}
  {2025}{\natexlab{b}}),\ \Eprint {https://arxiv.org/abs/2502.12364}
  {arXiv:2502.12364} \BibitemShut {NoStop}%
\bibitem [{\citenamefont {Cassidy}\ \emph
  {et~al.}(2024{\natexlab{b}})\citenamefont {Cassidy}, \citenamefont
  {Hofierka}, \citenamefont {Cunningham},\ and\ \citenamefont
  {Green}}]{Cassidy2024_2}%
  \BibitemOpen
  \bibfield  {author} {\bibinfo {author} {\bibfnamefont {J.~P.}\ \bibnamefont
  {Cassidy}}, \bibinfo {author} {\bibfnamefont {J.}~\bibnamefont {Hofierka}},
  \bibinfo {author} {\bibfnamefont {B.}~\bibnamefont {Cunningham}},\ and\
  \bibinfo {author} {\bibfnamefont {D.~G.}\ \bibnamefont {Green}},\ }\bibfield
  {title} {\bibinfo {title} {Many-body theory calculations of positronic-bonded
  molecular dianions},\ }\href {https://doi.org/10.1063/5.0188719} {\bibfield
  {journal} {\bibinfo  {journal} {J.~Chem.~Phys.}\ }\textbf {\bibinfo {volume}
  {160}},\ \bibinfo {pages} {084304} (\bibinfo {year}
  {2024}{\natexlab{b}})}\BibitemShut {NoStop}%
\bibitem [{\citenamefont {Shao}\ \emph {et~al.}(2016)\citenamefont {Shao},
  \citenamefont {{da Jornada}}, \citenamefont {Yang}, \citenamefont
  {Deslippe},\ and\ \citenamefont {Louie}}]{Shao:2016}%
  \BibitemOpen
  \bibfield  {author} {\bibinfo {author} {\bibfnamefont {M.}~\bibnamefont
  {Shao}}, \bibinfo {author} {\bibfnamefont {F.~H.}\ \bibnamefont {{da
  Jornada}}}, \bibinfo {author} {\bibfnamefont {C.}~\bibnamefont {Yang}},
  \bibinfo {author} {\bibfnamefont {J.}~\bibnamefont {Deslippe}},\ and\
  \bibinfo {author} {\bibfnamefont {S.~G.}\ \bibnamefont {Louie}},\ }\bibfield
  {title} {\bibinfo {title} {Structure preserving parallel algorithms for
  solving the {B}ethe–{S}alpeter eigenvalue problem},\ }\href
  {https://doi.org/https://doi.org/10.1016/j.laa.2015.09.036} {\bibfield
  {journal} {\bibinfo  {journal} {Linear Algebra Appl.}\ }\textbf {\bibinfo
  {volume} {488}},\ \bibinfo {pages} {148} (\bibinfo {year}
  {2016})}\BibitemShut {NoStop}%
\bibitem [{\citenamefont {Riso}\ \emph {et~al.}(2026)\citenamefont {Riso},
  \citenamefont {Trabski}, \citenamefont {Rossi}, \citenamefont {Green},\ and\
  \citenamefont {Koch}}]{Rosario2026}%
  \BibitemOpen
  \bibfield  {author} {\bibinfo {author} {\bibfnamefont {R.~R.}\ \bibnamefont
  {Riso}}, \bibinfo {author} {\bibfnamefont {J.~H.~M.}\ \bibnamefont
  {Trabski}}, \bibinfo {author} {\bibfnamefont {F.}~\bibnamefont {Rossi}},
  \bibinfo {author} {\bibfnamefont {D.~G.}\ \bibnamefont {Green}},\ and\
  \bibinfo {author} {\bibfnamefont {H.}~\bibnamefont {Koch}},\ }\bibfield
  {title} {\bibinfo {title} {Coupled-cluster theory for positron binding in
  anions and polyatomic molecules},\ }\href {https://doi.org/10.1063/5.0334768}
  {\bibfield  {journal} {\bibinfo  {journal} {J. Chem. Phys.}\ }\textbf
  {\bibinfo {volume} {165}},\ \bibinfo {pages} {034121} (\bibinfo {year}
  {2026})}\BibitemShut {NoStop}%
\bibitem [{\citenamefont {Prokof'ev}\ and\ \citenamefont
  {Svistunov}(1998)}]{Prokofev1998}%
  \BibitemOpen
  \bibfield  {author} {\bibinfo {author} {\bibfnamefont {N.~V.}\ \bibnamefont
  {Prokof'ev}}\ and\ \bibinfo {author} {\bibfnamefont {B.~V.}\ \bibnamefont
  {Svistunov}},\ }\bibfield  {title} {\bibinfo {title} {Polaron problem by
  {D}iagrammatic {Q}uantum {M}onte {C}arlo},\ }\href
  {https://doi.org/10.1103/physrevlett.81.2514} {\bibfield  {journal} {\bibinfo
   {journal} {Physical Review Letters}\ }\textbf {\bibinfo {volume} {81}},\
  \bibinfo {pages} {2514} (\bibinfo {year} {1998})}\BibitemShut {NoStop}%
\bibitem [{\citenamefont {Van~Houcke}\ \emph {et~al.}(2010)\citenamefont
  {Van~Houcke}, \citenamefont {Kozik}, \citenamefont {Prokof'ev},\ and\
  \citenamefont {Svistunov}}]{VanHoucke2010}%
  \BibitemOpen
  \bibfield  {author} {\bibinfo {author} {\bibfnamefont {K.}~\bibnamefont
  {Van~Houcke}}, \bibinfo {author} {\bibfnamefont {E.}~\bibnamefont {Kozik}},
  \bibinfo {author} {\bibfnamefont {N.}~\bibnamefont {Prokof'ev}},\ and\
  \bibinfo {author} {\bibfnamefont {B.}~\bibnamefont {Svistunov}},\ }\bibfield
  {title} {\bibinfo {title} {Diagrammatic {M}onte {C}arlo},\ }\href
  {https://doi.org/10.1016/j.phpro.2010.09.034} {\bibfield  {journal} {\bibinfo
   {journal} {Physics Procedia}\ }\textbf {\bibinfo {volume} {6}},\ \bibinfo
  {pages} {95} (\bibinfo {year} {2010})}\BibitemShut {NoStop}%
\bibitem [{\citenamefont {Van~Houcke}\ \emph {et~al.}(2012)\citenamefont
  {Van~Houcke}, \citenamefont {Werner}, \citenamefont {Kozik}, \citenamefont
  {Prokof'ev}, \citenamefont {Svistunov}, \citenamefont {Ku}, \citenamefont
  {Sommer}, \citenamefont {Cheuk}, \citenamefont {Schirotzek},\ and\
  \citenamefont {Zwierlein}}]{VanHoucke2012}%
  \BibitemOpen
  \bibfield  {author} {\bibinfo {author} {\bibfnamefont {K.}~\bibnamefont
  {Van~Houcke}}, \bibinfo {author} {\bibfnamefont {F.}~\bibnamefont {Werner}},
  \bibinfo {author} {\bibfnamefont {E.}~\bibnamefont {Kozik}}, \bibinfo
  {author} {\bibfnamefont {N.}~\bibnamefont {Prokof'ev}}, \bibinfo {author}
  {\bibfnamefont {B.}~\bibnamefont {Svistunov}}, \bibinfo {author}
  {\bibfnamefont {M.~J.~H.}\ \bibnamefont {Ku}}, \bibinfo {author}
  {\bibfnamefont {A.~T.}\ \bibnamefont {Sommer}}, \bibinfo {author}
  {\bibfnamefont {L.~W.}\ \bibnamefont {Cheuk}}, \bibinfo {author}
  {\bibfnamefont {A.}~\bibnamefont {Schirotzek}},\ and\ \bibinfo {author}
  {\bibfnamefont {M.~W.}\ \bibnamefont {Zwierlein}},\ }\bibfield  {title}
  {\bibinfo {title} {Feynman diagrams versus {F}ermi-gas {F}eynman emulator},\
  }\href {https://doi.org/10.1038/nphys2273} {\bibfield  {journal} {\bibinfo
  {journal} {Nature Physics}\ }\textbf {\bibinfo {volume} {8}},\ \bibinfo
  {pages} {366} (\bibinfo {year} {2012})}\BibitemShut {NoStop}%
\bibitem [{\citenamefont {Chen}\ and\ \citenamefont {Haule}(2019)}]{Chen2019}%
  \BibitemOpen
  \bibfield  {author} {\bibinfo {author} {\bibfnamefont {K.}~\bibnamefont
  {Chen}}\ and\ \bibinfo {author} {\bibfnamefont {K.}~\bibnamefont {Haule}},\
  }\bibfield  {title} {\bibinfo {title} {A combined variational and
  diagrammatic quantum {M}onte {C}arlo approach to the many-electron problem},\
  }\href {https://doi.org/10.1038/s41467-019-11708-6} {\bibfield  {journal}
  {\bibinfo  {journal} {Nature Comms.}\ }\textbf {\bibinfo {volume} {10}},\
  \bibinfo {pages} {3725} (\bibinfo {year} {2019})}\BibitemShut {NoStop}%
\bibitem [{\citenamefont {{\v{S}}imkovic}\ and\ \citenamefont
  {Rossi}(2021)}]{Fedor2021}%
  \BibitemOpen
  \bibfield  {author} {\bibinfo {author} {\bibfnamefont {F.}~\bibnamefont
  {{\v{S}}imkovic}}\ and\ \bibinfo {author} {\bibfnamefont {R.}~\bibnamefont
  {Rossi}},\ }\bibfield  {title} {\bibinfo {title} {Many-configuration
  markov-chain {M}onte {C}arlo}\ }(\bibinfo {year} {2021})\ \Eprint
  {https://arxiv.org/abs/2102.05613} {arXiv:2102.05613} \BibitemShut {NoStop}%
\bibitem [{\citenamefont {Azadi}\ \emph {et~al.}(2022)\citenamefont {Azadi},
  \citenamefont {Davydov},\ and\ \citenamefont {Kozik}}]{Azadi2022}%
  \BibitemOpen
  \bibfield  {author} {\bibinfo {author} {\bibfnamefont {S.}~\bibnamefont
  {Azadi}}, \bibinfo {author} {\bibfnamefont {A.}~\bibnamefont {Davydov}},\
  and\ \bibinfo {author} {\bibfnamefont {E.}~\bibnamefont {Kozik}},\ }\bibfield
   {title} {\bibinfo {title} {{$GW$} space-time method: Energy band gap of
  solid hydrogen},\ }\href {https://doi.org/10.1103/PhysRevB.105.155136}
  {\bibfield  {journal} {\bibinfo  {journal} {Phys. Rev. B}\ }\textbf {\bibinfo
  {volume} {105}},\ \bibinfo {pages} {155136} (\bibinfo {year}
  {2022})}\BibitemShut {NoStop}%
\bibitem [{\citenamefont {Bighin}\ \emph {et~al.}(2023)\citenamefont {Bighin},
  \citenamefont {Ho}, \citenamefont {Lemeshko},\ and\ \citenamefont
  {Tscherbul}}]{Bighin2023}%
  \BibitemOpen
  \bibfield  {author} {\bibinfo {author} {\bibfnamefont {G.}~\bibnamefont
  {Bighin}}, \bibinfo {author} {\bibfnamefont {Q.~P.}\ \bibnamefont {Ho}},
  \bibinfo {author} {\bibfnamefont {M.}~\bibnamefont {Lemeshko}},\ and\
  \bibinfo {author} {\bibfnamefont {T.~V.}\ \bibnamefont {Tscherbul}},\
  }\bibfield  {title} {\bibinfo {title} {Diagrammatic {M}onte {C}arlo for
  electronic correlation in molecules: High-order many-body perturbation theory
  with low scaling},\ }\href {https://doi.org/10.1103/physrevb.108.045115}
  {\bibfield  {journal} {\bibinfo  {journal} {Physical Review B}\ }\textbf
  {\bibinfo {volume} {108}},\ \bibinfo {pages} {045115} (\bibinfo {year}
  {2023})}\BibitemShut {NoStop}%
\bibitem [{\citenamefont {Sturt}\ and\ \citenamefont {Kozik}(2024)}]{John2024}%
  \BibitemOpen
  \bibfield  {author} {\bibinfo {author} {\bibfnamefont {J.}~\bibnamefont
  {Sturt}}\ and\ \bibinfo {author} {\bibfnamefont {E.}~\bibnamefont {Kozik}},\
  }\bibfield  {title} {\bibinfo {title} {Exploiting parallelism for fast
  {F}eynman diagrammatics}\ }(\bibinfo {year} {2024})\ \bibinfo {note}
  {arXiv:2502.10327},\ \Eprint {https://arxiv.org/abs/2501.00675} {2501.00675}
  \BibitemShut {NoStop}%
\bibitem [{\citenamefont {Vanhoecke}\ and\ \citenamefont
  {Schir{\`o}}(2024)}]{Vanhoecke2024}%
  \BibitemOpen
  \bibfield  {author} {\bibinfo {author} {\bibfnamefont {M.}~\bibnamefont
  {Vanhoecke}}\ and\ \bibinfo {author} {\bibfnamefont {M.}~\bibnamefont
  {Schir{\`o}}},\ }\bibfield  {title} {\bibinfo {title} {Diagrammatic {M}onte
  {C}arlo for dissipative quantum impurity models},\ }\href
  {https://doi.org/10.1103/PhysRevB.109.125125} {\bibfield  {journal} {\bibinfo
   {journal} {Phys. Rev. B}\ }\textbf {\bibinfo {volume} {109}},\ \bibinfo
  {pages} {125125} (\bibinfo {year} {2024})}\BibitemShut {NoStop}%
\bibitem [{\citenamefont {Brolli}\ \emph
  {et~al.}(2025{\natexlab{a}})\citenamefont {Brolli}, \citenamefont
  {Barbieri},\ and\ \citenamefont {Vigezzi}}]{Stefano2025}%
  \BibitemOpen
  \bibfield  {author} {\bibinfo {author} {\bibfnamefont {S.}~\bibnamefont
  {Brolli}}, \bibinfo {author} {\bibfnamefont {C.}~\bibnamefont {Barbieri}},\
  and\ \bibinfo {author} {\bibfnamefont {E.}~\bibnamefont {Vigezzi}},\
  }\bibfield  {title} {\bibinfo {title} {Diagrammatic {M}onte {C}arlo for
  finite systems at zero temperature},\ }\href
  {https://doi.org/10.1103/PhysRevLett.134.182502} {\bibfield  {journal}
  {\bibinfo  {journal} {Phys. Rev. Lett.}\ }\textbf {\bibinfo {volume} {134}},\
  \bibinfo {pages} {182502} (\bibinfo {year} {2025}{\natexlab{a}})}\BibitemShut
  {NoStop}%
\bibitem [{\citenamefont {Luo}\ \emph {et~al.}(2025)\citenamefont {Luo},
  \citenamefont {Park},\ and\ \citenamefont {Bernardi}}]{Luo2025}%
  \BibitemOpen
  \bibfield  {author} {\bibinfo {author} {\bibfnamefont {Y.}~\bibnamefont
  {Luo}}, \bibinfo {author} {\bibfnamefont {J.}~\bibnamefont {Park}},\ and\
  \bibinfo {author} {\bibfnamefont {M.}~\bibnamefont {Bernardi}},\ }\bibfield
  {title} {\bibinfo {title} {First-principles diagrammatic {M}onte {C}arlo for
  electron--phonon interactions and polaron},\ }\href
  {https://doi.org/10.1038/s41567-025-02954-1} {\bibfield  {journal} {\bibinfo
  {journal} {Nature Phys.}\ }\textbf {\bibinfo {volume} {21}},\ \bibinfo
  {pages} {1275} (\bibinfo {year} {2025})}\BibitemShut {NoStop}%
\bibitem [{\citenamefont {Brolli}\ \emph
  {et~al.}(2025{\natexlab{b}})\citenamefont {Brolli}, \citenamefont
  {Barbieri},\ and\ \citenamefont {Vigezzi}}]{Brolli}%
  \BibitemOpen
  \bibfield  {author} {\bibinfo {author} {\bibfnamefont {S.}~\bibnamefont
  {Brolli}}, \bibinfo {author} {\bibfnamefont {C.}~\bibnamefont {Barbieri}},\
  and\ \bibinfo {author} {\bibfnamefont {E.}~\bibnamefont {Vigezzi}},\
  }\bibfield  {title} {\bibinfo {title} {Diagrammatic monte carlo for finite
  systems at zero temperature},\ }\href
  {https://doi.org/10.1103/PhysRevLett.134.182502} {\bibfield  {journal}
  {\bibinfo  {journal} {Phys. Rev. Lett.}\ }\textbf {\bibinfo {volume} {134}},\
  \bibinfo {pages} {182502} (\bibinfo {year} {2025}{\natexlab{b}})}\BibitemShut
  {NoStop}%
\bibitem [{Note3()}]{Note3}%
  \BibitemOpen
  \bibinfo {note} {Since the self-energy matrix is symmetric, only the upper
  triangle ($i \leq f$) is sampled, with the calculation parallelised over
  unique $(i,f)$ pairs.}\BibitemShut {Stop}%
\bibitem [{\citenamefont {Patterson}(2020)}]{DF1}%
  \BibitemOpen
  \bibfield  {author} {\bibinfo {author} {\bibfnamefont {C.~H.}\ \bibnamefont
  {Patterson}},\ }\bibfield  {title} {\bibinfo {title} {Density fitting in
  periodic systems: Application to {TDHF} in diamond and oxides},\ }\href
  {https://doi.org/10.1063/5.0014106} {\bibfield  {journal} {\bibinfo
  {journal} {J. Chem. Phys.}\ }\textbf {\bibinfo {volume} {153}},\ \bibinfo
  {pages} {064107} (\bibinfo {year} {2020})}\BibitemShut {NoStop}%
\bibitem [{Note4()}]{Note4}%
  \BibitemOpen
  \bibinfo {note} {The proposal probabilities are user-configurable. For
  example, selecting only the positron--electron interaction with all others
  set to zero yields the $\Gamma $ ladder series. When multiple interaction
  types are active, their proposal probabilities must be equal to satisfy
  detailed balance.}\BibitemShut {Stop}%
\bibitem [{\citenamefont {K{\"o}rle}(1970)}]{Korle1970}%
  \BibitemOpen
  \bibfield  {author} {\bibinfo {author} {\bibfnamefont {H.-H.}\ \bibnamefont
  {K{\"o}rle}},\ }\bibfield  {title} {\bibinfo {title} {On absolute summability
  by {R}iesz and generalized {C}es\`aro means. {I}},\ }\href
  {https://doi.org/10.4153/CJM-1970-026-6} {\bibfield  {journal} {\bibinfo
  {journal} {Canadian J. Math.}\ }\textbf {\bibinfo {volume} {22}},\ \bibinfo
  {pages} {202} (\bibinfo {year} {1970})}\BibitemShut {NoStop}%
\bibitem [{\citenamefont {Green}\ \emph {et~al.}(2018)\citenamefont {Green},
  \citenamefont {Swann},\ and\ \citenamefont {Gribakin}}]{Green2018}%
  \BibitemOpen
  \bibfield  {author} {\bibinfo {author} {\bibfnamefont {D.~G.}\ \bibnamefont
  {Green}}, \bibinfo {author} {\bibfnamefont {A.~R.}\ \bibnamefont {Swann}},\
  and\ \bibinfo {author} {\bibfnamefont {G.~F.}\ \bibnamefont {Gribakin}},\
  }\bibfield  {title} {\bibinfo {title} {Many-body theory for positronium-atom
  interactions},\ }\href {https://doi.org/10.1103/PhysRevLett.120.183402}
  {\bibfield  {journal} {\bibinfo  {journal} {Phys. Rev. Lett.}\ }\textbf
  {\bibinfo {volume} {120}},\ \bibinfo {pages} {183402} (\bibinfo {year}
  {2018})}\BibitemShut {NoStop}%
\bibitem [{\citenamefont {Swann}\ \emph {et~al.}(2023)\citenamefont {Swann},
  \citenamefont {Green},\ and\ \citenamefont {Gribakin}}]{Swann2023}%
  \BibitemOpen
  \bibfield  {author} {\bibinfo {author} {\bibfnamefont {A.~R.}\ \bibnamefont
  {Swann}}, \bibinfo {author} {\bibfnamefont {D.~G.}\ \bibnamefont {Green}},\
  and\ \bibinfo {author} {\bibfnamefont {G.~F.}\ \bibnamefont {Gribakin}},\
  }\bibfield  {title} {\bibinfo {title} {Many-body theory of positronium
  scattering and pickoff annihilation in noble-gas atoms},\ }\href
  {https://doi.org/10.1103/PhysRevA.107.042802} {\bibfield  {journal} {\bibinfo
   {journal} {Phys. Rev. A}\ }\textbf {\bibinfo {volume} {107}},\ \bibinfo
  {pages} {042802} (\bibinfo {year} {2023})}\BibitemShut {NoStop}%
\end{thebibliography}
%

\end{document}